\documentclass[aps, prd, twocolumn, floats, floatfix, superscriptaddress, nofootinbib]{revtex4}

\usepackage[utf8]{inputenc}
\usepackage{amsfonts,amsmath,amssymb,amsthm,mathtools}
\usepackage[dvipsnames]{xcolor}
\usepackage[sort&compress]{natbib} 
\usepackage[linkcolor=blue,citecolor=blue,colorlinks=true]{hyperref}
\usepackage{graphicx}
\graphicspath{{figures/}} 

\usepackage{latexsym}
\usepackage[title]{appendix}
\usepackage{aas_macros}
\usepackage{siunitx}
\usepackage[normalem]{ulem}

\newcommand{\A}{{\mathcal A}}
\newcommand{\B}{{\mathcal B}}
\newcommand{\C}{{\mathcal C}}
\def\pert{\varepsilon}
\def\manifold{\mathcal{M}}

\def\stax{\tiny{\mbox{STAX}}}

\def\gback{g}
\def\fpt{K_1}
\def\spt{K_2}
\def\tidalpt{K_1^T}

\def\tOmega{\tilde{\Omega}}

\def\defor{\Xi}

\def\rstar{R}
\def\ro{\rstar}
\def\Qone{Q_1}
\def\comega{c_\omega^2}
\def\crho{c_\rho^2}
\def\csigma{c_\sigma^2}
\def\kn{k_n}
\def\kp{k_p}
\def\ms{m_\star\lvert_0}
\def\mds{m_\star^2\lvert_0}
\def\me{m_e}

\def\pert{\varepsilon}
\def\manifold{\mathcal{M}}

\def\diff{{\rm{d}}}

\def\constK{C}

\def\compact{C_o}

\def\press{\mathcal{P}}

\def\normal{\mathfrak{n}}

\def\gauge{\Gamma}
\def\hgauge{\Upsilon}
\def\family{\tilde}

\begin{document}

\title{
\boldmath Revisiting the  $I$-Love-$Q$ relations for superfluid neutron stars\unboldmath\\
}

\author{Eneko Aranguren}
\email{eneko.aranguren@ehu.eus}
\affiliation{Fisika Saila, University of the Basque Country UPV/EHU, 48080 Bilbao, Basque Country}
\author{Jos\'e A.~Font}
\email{j.antonio.font@uv.es}
\affiliation{Departamento de Astronom\'ia y Astrof\'isica, Universitat de Val\`encia, Dr. Moliner 50, 46100, Burjassot (Valencia), Spain}
\affiliation{Observatori Astron\`omic, Universitat de Val\`encia, Catedr\'atico Jos\'e Beltr\'an 2, 46980, Paterna, Spain}
\author{Nicolas Sanchis-Gual}
\email{nicolas.sanchis@uv.es}
\affiliation{Departamento de Astronom\'ia y Astrof\'isica, Universitat de Val\`encia, Dr. Moliner 50, 46100, Burjassot (Valencia), Spain}
 \author{Ra\"ul Vera}
\email{raul.vera@ehu.eus}
\affiliation{Fisika Saila, University of the Basque Country UPV/EHU, 48080 Bilbao, Basque Country}

\begin{abstract}
We study the tidal problem and the resulting $I$-Love-$Q$ approximate universal relations for rotating superfluid neutron stars in the Hartle-Thorne formalism.
Superfluid stars are described in this work by means of a two-fluid model consisting of superfluid neutrons and all other charged constituents. We employ a stationary and axisymmetric perturbation scheme to second order around a static and spherically symmetric background.
Recently, we used this scheme to study isolated rotating superfluid stars. In this paper it is applied
to analyze the axially symmetric sector of the tidal problem in a binary system.
We show that a consistent use of perturbative matching theory amends the original 
two-fluid formalism for the tidal problem
to account for the possible nonzero value of the energy density
at the boundary of the star. This is exemplified by building numerically different stellar models spanning three
equations of state.
Significant departures from universality are found when the correct matching relations
are not taken into account.
We also present an augmented set of universal relations for superfluid neutron stars which includes the contribution to the total mass of the star at second order, $\delta M$.
Therefore, our results  complete  the set of universal relations for rotating superfluid stars, generalizing our previous findings in the perfect fluid case.
\end{abstract}

\maketitle

\section{Introduction}

The analysis of the inspiral gravitational-wave signal emitted during a binary neutron star (BNS) coalescence provides information on the internal structure of neutron stars and on the supranuclear equation of state (EOS). In a BNS system, the tidal field of the companion induces a mass-quadrupole moment and accelerates the coalescence. The ratio of the induced quadrupole moment to the external tidal field is proportional to the tidal Love number of the star, $k_2$, or to the tidal deformability $\lambda_2=(2/3)k_2[(c^2/G)(R/M)]^5$, where $R$ and $M$ refer to the radius and mass of the star. The strength of tidal interactions increases rapidly during the final tens of gravitational-wave inspiral cycles before merger, making their effects potentially measurable~\cite{Hinderer:2010,Damour:2012,DelPozzo:2013,Agathos:2015,Harry:2018}. This has been put into practice in the analysis of GW170817 and GW190425, the first two (and, so far, only) BNS systems detected by the LIGO-Virgo-KAGRA Collaboration~\cite{LIGOScientific:2017vwq,LIGOScientific:2018cki,LIGOScientific:2018hze,GW190425}. The tidal deformability of these systems was measured using EOS-insensitive relations between the moment of inertia $I$, the tidal deformability  $\lambda_2$ (or the Love number $k_2$) and the spin-induced quadrupole moment $Q$, known as $I$-Love-$Q$ relations~\cite{Yagi:2013,Yagi:2013awa}. In the case of GW170817, the observational constraints on the  tidal deformation of the binary components allowed to rule out some of the stiffest supranuclear EOS models.
 
 The most basic theoretical treatment of the tidal problem in a binary system~\cite{Hinderer:2008,Hinderer2009} fits in the Hartle-Thorne scheme (HT hereafter) \cite{hartle1967,hartle2}, a pioneer proposal that provides a perturbative framework in General Relativity to describe the equilibrium configuration of a compact and isolated perfect fluid body around a static and spherically symmetric configuration, up to second order.
Within the HT scheme the tidal problem can be solved in the regime of stationary and axial perturbations (see~\cite{Damour:2009vw} and references therein).
 For this problem, the $I$-Love-$Q$ relations  found in~\cite{Yagi:2013} were first seen to split into two categories, one valid for ordinary neutron stars and another one for quark stars, the latter characterized by the presence of a nonvanishing energy density at the boundary of the star. Shortly after, 
 \cite{Yagi:2014} amended the results of~\cite{Yagi:2013} by considering
 a term used by~\cite{Hinderer:2010} {[see Eq.~(15) in the latter reference]}
 to account for a possible nonzero value of the energy density at the stellar boundary when computing the Love number. This, in turn, justified
 the result obtained by~\cite{Damour:2009vw}
 for the limiting case of homogeneous stars. The correction reported by~\cite{Yagi:2014} provided universal $I$-Love-$Q$ relations regardless of the EOS type.
 
 The proof that Eq.~(15) in \cite{Hinderer:2010} is indeed the correct expression for the Love number was reported in~\cite{reina-sanchis-vera-font2017} building on the amendment to the original HT scheme provided in \cite{ReinaVera2015}.
 We recall that the original HT scheme implicitly assumes that \emph{all} functions describing the perturbations are continuous everywhere, in particular at the boundary of the star. Apart from
 providing the needed results to put the HT scheme on firm grounds,
 the main
 point of the amendment
 was to prove the inconsistency of this assumption because,
 although most of the interior and exterior parts
 of the functions must indeed share the same
 value at the boundary,
 some of the functions do
 present a jump that is proportional
 to the value of the energy density there.

 The rigorous support and partial correction to the original HT scheme reported in \cite{ReinaVera2015} (see also \cite{MRV1,MRV2})
 is obtained by producing
 an initial framework resorting to perturbation
 theory in purely geometric terms. On top of that, the
 equations for the matter content at the stellar interior are to be imposed. Reference \cite{ReinaVera2015} focused on 
 perfect fluid stars (with barotropic EOS)
 finding
 that the discontinuity of one perturbation function due to the nonvanishing
 of the energy density at the boundary affects the computation
 of the contribution to the mass of the star at second order, $\delta M$
 (for a given fixed central pressure).
 This was first used in \cite{Reina:2015jia} to revisit 
 the seminal work of \cite{chandra_miller} on homogeneous rotating
 stars, and the significant correction to the total mass was underlined. Second, the correction in the computation
of the mass was used in~\cite{reina-sanchis-vera-font2017} to
show that $I$-Love-$Q$ EOS-insensitive relations also apply to $\delta M$,
thus extending the universality to a family of
four parameters, $I$-Love-$Q$-$\delta M$.

The original HT model was also adapted in~\cite{Andersson_Comer_2001} to describe slowly rotating, {\it superfluid} neutron stars, building on a two-fluid formalism introduced by \cite{Langlois:1998,comer1999}. This adaptation, however, inherited the incorrect (implicit) assumptions from the original HT scheme regarding the continuity of the perturbation functions at the stellar surface.
This has been recently fixed in~\cite{Aranguren:2022} where we have used the geometrical perturbation scheme of \cite{ReinaVera2015} (see also \cite{MRV2})
to amend the two-fluid formalism of isolated rotating superfluid stars.

Despite the fact that the results
in~\cite{ReinaVera2015,reina-sanchis-vera-font2017}
provide the perturbation formalism for the tidal problem with a geometrical justification to correctly compute the tidal number
leading to the universality of the $I$-Love-$Q$ relations, those works have been
overlooked by several subsequent studies.
In particular, the two-fluid model 
has been also used by \cite{Yeung:2021} to study the $I$-Love-$Q$ relations for superfluid neutron stars
imposing the continuity of all functions (and some derivatives)
without justification to compute the tidal deformability.
We note that, in principle, one cannot resort to any known explicit result or correct expression for the Love number, since those apply to the perfect fluid case.

The aim of the present paper is to explore the tidal problem
and the approximate universal relations for superfluid neutron stars, revisiting the results of \cite{Yeung:2021}
using the corrected HT scheme we started developing in our previous work \cite{Aranguren:2022}. As in~\cite{Andersson_Comer_2001,Char2018,Yeung:2021,Aranguren:2022} we describe superfluid stars by a simple two-fluid model which accounts for superfluid neutrons and all other constituents. Using a toy-model EOS for which the number densities of the two constituents do not vanish at the boundary of the star, we showed in \cite{Aranguren:2022} that the corrections to the HT formalism do impact the structure of rotating superfluid neutron stars in a significant way. In this paper we demonstrate that the study of the tidal problem for superfluid stars is also affected by the same continuity issues. Therefore, although we check
that the EOSs used in \cite{Yeung:2021, Char2018}
do not present those
issues due to the vanishing of the relevant physical quantities at
the boundary\footnote{Unfortunately, we find no explicit mention on the behavior of the fluid quantities at the boundary in those works.}, the correction of the HT formalism we report here needs to be considered for general-purpose (i.e.~EOS insensitive) computations of the tidal problem in a binary system.

The structure of this paper is as follows: In Sec. \ref{sec:two-fluid} we
briefly recall the two-fluid formalism and the construction of the global interior/exterior configuration.  In Secs. \ref{sec:perturbationscheme} and \ref{sec:background} we briefly describe the perturbation scheme for a two-fluid model and we develop the background configuration of superfluid neutron stars. Thus, these sections lay the groundwork for the notation that will be employed later on. Next, Secs. \ref{sec:firstorder} and \ref{sec:secondorder}  describe the first and second order problems, respectively. Once the general setup has been constructed, Sec. \ref{sec:tidal} addresses the tidal problem and the Love numbers obtained therein. Correspondingly, Sec. \ref{sec:results} presents our results to test the universality of our $I$-Love-$Q$-$\delta M$ relations, for a variety of physically motivated EOS, as well as a toy model. Our conclusions are summarized in Sec. \ref{sec:conclusions}.
Unless otherwise stated (see, for example, Appendix \ref{app:units}), we will be using units such that $G=c=1$.

\section{Two-fluid model}\label{sec:two-fluid}
 In the two-fluid formalism, as
 originally developed in \cite{Carter_Lecture_Notes:1989, Langlois:1998}
 (see also \cite{comer1999, Andersson_Comer_2001}),
the flow of neutrons and protons
are described, respectively, by two vectors
\begin{equation*}
    n^{\alpha}=nu^{\alpha},\quad p^{\alpha}=pv^{\alpha},
\end{equation*}
where $u^\alpha$ and $v^\alpha$ are two unit timelike vectors
and $n$ and $p$ are the neutron and proton number densities.
The coupling of the neutrons and protons is described
by the quantity $x^2:=-p_\alpha n^\alpha$.
The EOS of the whole system is provided by
specifying a master function
\[
  \Lambda = \Lambda(n^2,p^2,x^2),
\]
that depends on three arguments.

In terms of the auxiliary functions
\begin{align*}
&\A:=-\frac{\partial\Lambda(n^2,p^2,x^2)}{\partial x^2},\quad
\B:=-2\frac{\partial\Lambda(n^2,p^2,x^2)}{\partial n^2},\\
&\C:=-2\frac{\partial\Lambda(n^2,p^2,x^2)}{\partial p^2},
\end{align*}
the 1-forms
\begin{equation*}
    \mu_{\alpha}:=\B n_{\alpha}+\A p_{\alpha},\qquad\chi_{\alpha}:=\C p_{\alpha}+\A n_{\alpha},
\end{equation*}
are the dynamically and thermodynamically conjugates to $n^\alpha$ and $p^\alpha$,
respectively.
The energy-momentum tensor of the fluid is then given by
\begin{equation}\label{eq:Tmunu}
  T^\alpha{}_{\beta}=\Psi \delta^\alpha_{\beta}+p^\alpha \chi_\beta+n^\alpha \mu_\beta,
\end{equation}
where
\begin{equation}\label{eq:generalizedpressure}
    \Psi:=\Lambda-n^{\alpha}\mu_{\alpha}-p^{\alpha}\chi_{\alpha},
\end{equation}
acts as a generalized pressure.

The equations of motion are given by the conservation equations
\begin{equation}
  \nabla_\alpha n^\alpha=0,\quad \nabla_\alpha p^\alpha=0,\label{eq:conservation}
\end{equation}
plus the Euler equations
\begin{equation}\label{eq:euler}
  n^\alpha (\nabla_{\alpha} \mu_{\beta}-\nabla_{\beta} \mu_{\alpha})=0,\quad
    p^\alpha (\nabla_{\alpha} \chi_{\beta}-\nabla_{\beta} \chi_{\alpha})=0.
\end{equation}
Equations \eqref{eq:conservation} and \eqref{eq:euler} imply $\nabla^\alpha T_{\alpha\beta}=0$.

The two problems at hand will be framed
in a stationary and axially symmetric setting (describing the perturbations)
over a static and spherically symmetric background configuration.
As customary, we use spherical coordinates $\{t,r,\theta,\phi\}$
arranged so that the timelike and axial (spacelike) Killing vector fields
of the whole setting read $\partial_t$ and $\partial_\phi$ respectively.
Thus, the functions describing the stationary and axisymmetric
spacetime geometry $g_{\stax}$
and the fluids only depend on $r$ and $\theta$.

Moreover, if the fluids
are assumed to rotate around the axis so that there are no convective
motions, and the rotation is rigid, then
\begin{equation}\label{eq:uv}
  u\propto\left(\partial_t+\tOmega_n \partial_\phi\right),\quad
  v\propto\left(\partial_t+\tOmega_p \partial_\phi\right),
\end{equation}
for some constants $\tOmega_n$ and $\tOmega_p$, which represent the angular velocities of neutrons and protons, respectively.
In this case \eqref{eq:conservation} are automatically satisfied
and \eqref{eq:euler} are equivalent to
\begin{align}\label{eq:euler_const}
  \mu_c=-g_{\stax}(\partial_t+\tOmega_n\partial_\phi,\mu),\nonumber\\
  \chi_c=-g_{\stax}(\partial_t+\tOmega_p\partial_\phi,\chi),
\end{align}
for some constants $\mu_c$ and $\chi_c$.
We use $g_{\stax}(\cdot,\cdot)$ for the scalar product in the index-free notation.

\subsection{Global configuration: Vacuum exterior}\label{sec:global_setting}
The global model of the star consists of two spacetimes
$(\manifold^+,g_{\stax}^+)$ and $(\manifold^-,g_{\stax}^-)$ with timelike boundaries $\Sigma^+$
and $\Sigma^-$ which are pointwise identified $\Sigma\equiv \Sigma^+=\Sigma^-$,
to produce a joined  spacetime $(\manifold,g_{\stax})$
with $\manifold=\manifold^+\cup\manifold^-$,
and $g_{\stax}$ is $g^\pm_{\stax}$ on each region $\manifold^\pm$ accordingly.
The identification is required to be isometric, so that $\Sigma$ has an induced
metric $h$. This requires the well-known first matching (or junction) conditions,
$h:=h^+=h^-$, where $h^\pm$ are the induced metrics
of $\Sigma$ as embedded on $(\manifold^\pm,g^\pm_{\stax})$, respectively.
Then, $g_{\stax}$ can be extended continuously on $\manifold$.
To avoid a distributional Riemann tensor on $(\manifold,g_{\stax})$,
which is equivalent to avoid energy surface layers at the boundary of the star in General Relativity,
we must demand that the second fundamental forms (extrinsic curvatures) $\kappa^\pm$ of $\Sigma$ 
as embedded on $(\manifold^\pm,g^\pm_{\stax})$ agree. To sum up, the full matching conditions
require then that $h^+=h^-$ and $\kappa^+=\kappa^-$ hold on $\Sigma$.

We take the $+$ part to describe the interior of the star,
thus solving the two-fluid model problem,
and the $-$ part to describe the vacuum exterior.
The (history of the) surface of the star is provided by $\Sigma$.
For the two problems we are interested in, we assume that both spacetimes are stationary
and axisymmetric, so that the boundaries inherit the two symmetries \cite{Vera2002}.

The interior and exterior problems are then solved imposing ``regularity''
at the origin, and whatever conditions we want to impose on the exterior,
in addition to the relations on $\Sigma$ provided by the matching
conditions. 
In particular, the matching conditions also determine, in principle,
the surface of the star.
As shown in \cite{Aranguren:2022} for our stationary and axisymmetric
setting, the matching conditions imply
the continuity of $\Psi$ across $\Sigma$. Therefore
\[
  \Psi(r,\theta)=0
\]
determines the surface of the star $r=r(\theta)$ implicitly.
This condition is not sufficient, but it is the only one involving only the interior side,
as shown in~\cite{Vera:2003} (see also \cite{Mars:1998})
for the general stationary and axisymmetric setting.
The rest of the matching conditions provide the matching hypersurface
from the other side $\Sigma^-$ and relations between the
boundary data for the interior and exterior problems.

In the following we will denote by $[f]$, for any function $f$,
the difference of $f$ evaluated at both sides of the hypersurface
$\Sigma$, i.e. $[f](p):=f^+|_{p^+\in\Sigma^+}-f^-|_{p^-\in\Sigma^-}$
where $p(\in\Sigma)=p^+=p^-$ after the identification.
Moreover, if $[f]=0$ we will simply use $f$ when evaluated on either $\Sigma^+$ or $\Sigma^-$. All expressions in square brackets will denote their difference.  

\section{Perturbation scheme}\label{sec:perturbationscheme}
In this paper we focus on two problems, namely, the deformation
due to the rotation of an isolated star, and the axially symmetric sector of the (even-parity) tidal problem caused by a companion star.
The two issues are dealt with as two perturbative stationary and axisymmetric
problems over a static and spherically
symmetric configuration.

For the perturbative problems we use perturbation theory in metric theories 
of gravity, which is, in effect,
a gauge-field theory of symmetric tensors on a given background configuration at each order.
In particular, our work is
based on the concept of perturbation scheme, which includes the notion of classes
of gauges, that inherits some of the symmetries of the background.
We refer the reader to \cite{MRV1} for the detailed definitions
and the gauge fixing procedures involved in axially symmetric and axistationary
second order perturbations around spherical backgrounds.
To deal with the matching of the exterior and interior regions we use
the theory of perturbed matchings,
based on perturbations of hypersurfaces to second order \cite{Mars:2005},
particularized to the case of stationary and
axially symmetric perturbations in \cite{MRV2} (see also \cite{ReinaVera2015}). 

\subsection{Perturbation theory in rigidly rotating two-fluid stars}
The perturbative problem of the isolated rotating star modeled by a two-fluid
has been already dealt with in our previous reference \cite{Aranguren:2022}, revisiting and amending
the approach and results in \cite{Andersson_Comer_2001}.
However, for completeness we include here an outline of the whole procedure
because the tidal problem shares most part of the setting.
Thus, we follow the stationary and axisymmetric perturbative scheme to second
order around a static and spherically symmetric background
$(\manifold, g)$ as described in \cite{MRV1} (see also \cite{MRV2,ReinaVera2015}) based
on an abstract perturbation parameter $\pert$.
In short, we have a family of stationary and axisymmetric
spacetimes $(\manifold_\pert,\family g_\pert)$, where
$(\manifold_0,\family g_0)=(\manifold,g)$ is our static
and spherically symmetric background, together with a class of point
identifications $\gauge_\pert:\manifold\to \manifold_\pert$
(spacetime gauges), where $\gauge_0$ is the identity.
This class of gauges is, so far, only restricted to
inherit the stationarity and axial symmetry generated by $\partial_t$ and $\partial_\phi$
in the background as defined in \cite{MRV1}.

On each $(\manifold_\pert,\family g_\pert)$ we have defined the two-fluid model quantities,
and the equations they satisfy,
that depend on $\pert$. The metrics $\family g_\pert$ as well as
all the fluid quantities, and the corresponding equations,
are pulled back using $\gauge_\pert^*$ onto $(\manifold,g)$.

In particular, the procedure defines a family of metrics $g_\pert=\gauge_\pert^*(\family g_\pert)$
on $\manifold$.
The first order $K_1$ and second order $K_2$ perturbation
tensors are defined as the first and second order derivatives of $g_\pert$
with respect to $\pert$, evaluated at $\pert=0$.
As a result, the $\pert$-family of metrics
can be written as the usual
\[
g_\pert = \gback + \pert\fpt + \frac{1}{2}\pert^2\spt+ O(\pert^3).
\]
If we take, as explained above, spherical coordinates $\{t,r,\theta,\phi\}$
on  $(\manifold,g)$,
then the inheriting of the symmetries by the class of gauges $\gauge_\pert$
means that $\partial_t$ and $\partial_\phi$ are Killings of the whole family $g_\pert$.
Therefore,
just like $g$,
the perturbation tensors $K_1$ and $K_2$ do not depend on $t$ nor $\phi$.
Now, suitable gauge-fixing procedures can be used to simplify further
the forms of $K_1$ and $K_2$.

Similarly, for every two-fluid model quantity we have a corresponding $\pert$-family of
quantities defined on $\manifold$, and thence background, first and second order
corresponding quantities.
Explicitly, the number density of neutrons and protons
are decomposed to second order as (we follow the notation from \cite{Andersson_Comer_2001})
\begin{align}
    &n_\pert(r,\theta)=n_0(r)\left(1+\pert^2\eta(r,\theta)\right)+O(\pert^3),\\
    &p_\pert(r,\theta)=p_0(r)\left(1+\pert^2\Phi(r,\theta)\right)+O(\pert^3).
\end{align}
The fact that there is no contribution at first order
  is a consequence of the forms $K_1$ and $K_2$ take in
  the stationary and axisymmetric perturbative setting over
  a static and spherical background configuration.
  A rigorous account on this matters is made in \cite{MRV1,MRV2}
  for the perfect fluid case.
  For the purposes of this work we will assume the usual forms of the
  perturbation tensors and this decomposition for the two-fluid quantities,
  which is consistent,   from the beginning.

As in the perfect fluid case, where the same equation of state is assumed
for the whole (background and perturbations) configuration,
here one demands
$\Lambda_\pert(n_\pert^2,p_\pert^2,x_\pert^2)=\Lambda(n_\pert^2,p_\pert^2,x_\pert^2)$.
In the following we use
the notation $\Lambda_\pert:=\Lambda(n_\pert^2,p_\pert^2,x_\pert^2)$,
so that $\Lambda_0=\Lambda(n_0^2,p_0^2,x_0^2)$.
We will also use
$\Lambda_0(r):=\Lambda(n_0^2(r),p_0^2(r),x_0^2(r))$
and equivalently for $\Psi_0(r)$.
The flows $u_\pert$ and $v_\pert$ have the form of \eqref{eq:uv}
with some $\tOmega_{n\pert}$ and $\tOmega_{p\pert}$ (only dependency on $\pert$).
Since the background is static we have $\tOmega_{n 0}=\tOmega_{p 0}=0$.
On the other hand, it is (implicitly) assumed that after a redefinition of the
perturbation parameter to absorb the second order contributions in $\tOmega$'s,
we have\footnote{The fact that the redefinition of the perturbation parameter $\pert$
  to absorb the second order contribution to $\Omega$
  is consistent with the problem to second order should be proven
  after all the problem has been set.
  To our knowledge this has been only (rigorously) proven
  in the rigidly rotating perfect fluid case, in \cite{MRV1,MRV2}.}
\begin{align*}
  \tOmega_{n\pert}= \pert\Omega_n +O(\pert^3),\\
  \tOmega_{p\pert}= \pert\Omega_p +O(\pert^3),
\end{align*}
for some pair of constants $\Omega_n$ and $\Omega_p$.
The full form of the flows $u_\pert$ and $v_\pert$
as well as the set of $\pert$-families, $\{x_\pert,\mu^\alpha_\pert,\chi^\alpha_\pert,\Psi_\pert,\mu_{c\pert},\chi_{c \pert}\}$,
are then found using the expressions from Sec. \ref{sec:two-fluid},
taking into account that
\begin{align}
  &\A_\pert=-\frac{\partial\Lambda(n_\pert^2,p_\pert^2,x_\pert^2)}{\partial x_\pert^2},\quad\B_\pert=-2\frac{\partial\Lambda(n_\pert^2,p_\pert^2,x_\pert^2)}{\partial n_\pert^2},\nonumber\\
  &\C_\pert=-2\frac{\partial\Lambda(n_\pert^2,p_\pert^2,x_\pert^2)}{\partial p_\pert^2},\label{eq:ABC}
\end{align}
plus the tensors $K_1$ and $K_2$.
From those quantities we construct $T_\pert{}^\alpha{}_\beta$ using \eqref{eq:Tmunu}
accordingly.
The expressions for the rotating perturbation case (to second order)
are given in full in \cite{Aranguren:2022}.

Since the Einstein field equations hold on each $(\manifold_\pert,\family g_\pert)$,
the corresponding pullbacks onto $(\manifold,g)$ must also hold, and therefore
\begin{equation}\label{eq:Einstein_e}
  Ein(g_\pert)^\alpha{}_\beta=\varkappa T_\pert{}^\alpha{}_\beta,
\end{equation}
must be satisfied for all $\pert$, where $\varkappa=8\pi G/c^4$ and $Ein(g_\pert)$ is the Einstein
tensor computed from $g_\pert$.
The background equations are \eqref{eq:Einstein_e} evaluated
at $\pert=0$, while the first and second order Einstein equations correspond to
the first and second order derivatives with respect to $\pert$ evaluated at $\pert=0$
respectively.

Similarly, the Euler equations \eqref{eq:euler_const} apply for $g_{\stax}=g_\pert$
and all the quantities substituted by their $\pert$-counterparts on the right-hand side.
To use the notation of \cite{Andersson_Comer_2001,Aranguren:2022},
the $\pert$-families of constants $\mu_{c \pert}$ and $\chi_{c \pert}$ are explicitly
written as
\begin{align}
  \label{eq:const_euler_2}
  \mu_c{}_\pert=\mu_\infty\left(1+\pert^2 \gamma_n\right)+O(\pert^3),\nonumber\\
  \chi_c{}_\pert=\chi_\infty\left(1+\pert^2 \gamma_p\right)+O(\pert^3),
\end{align}
which define
the four constants $\mu_\infty(=\mu_c{}_0)$, $\chi_\infty(=\chi_c{}_0)$, $\gamma_n$ and $\gamma_p$.

We finish this section with a brief comment on the perturbation parameters.
Let us first stress that, apart from the boundary data needed to solve the background
configuration, the exact model only contains two free parameters.
These correspond to the rotating parameters $\tOmega_n$ and $\tOmega_p$.
In the perturbative approach we have instead three, namely
$\Omega_n$, $\Omega_p$ and $\pert$. The introduction of a spurious
parameter is a consequence of the scalability property of perturbation theory.
Computationally one chooses freely one of the three parameters, say $\Omega_p=1$. Then,
for a desired value of the relative rotation rate $\Delta:=\Omega_n/\Omega_p$ fixes $\Omega_n$
accordingly. After solving the problems, one finds the convenient measurable physical quantities and
uses the scalability property to fix the model to the data needed.

\subsection{Perturbed matching}
Let us be given a static and spherically symmetric background
global configuration $(\manifold,g)$,
composed by $(\manifold^+, g^+,\Sigma^+)$ and
$(\manifold^-, g^-,\Sigma^-)$ with identified boundaries $\Sigma:=\Sigma^+=\Sigma^-$,
and such that the matching conditions $h^+=h^-$ and $\kappa^+=\kappa^-$ hold on $\Sigma$.
Assume now that the global configuration setting described in Sec. \ref{sec:global_setting}
applies to a $\pert$-family
of spacetimes $(\manifold_\pert,\family g_\pert)$ such that $(\manifold,g)=(\manifold_0,\family g_0)$.
That is, we take $(\manifold_\pert,\family g_\pert)$,
for each $\pert$ around $0$,
to be composed by two spacetimes with boundary
$(\manifold^+_\pert,\family g^+_\pert,\family\Sigma_\pert^+)$ and
$(\manifold^-_\pert,\family g^-_\pert,\family\Sigma_\pert^-)$ so that $\family\Sigma_\pert:=\family\Sigma_\pert^+=\family\Sigma_\pert^-$ after some identification of points, and $\manifold_\pert^+\cap\manifold_\pert^-=\family\Sigma_\pert$.
The matching conditions $\family h_\pert^+=\family h_\pert^-$,
$\family \kappa_\pert^+=\family \kappa_\pert^-$  are satisfied on each $\family\Sigma_\pert$ by construction, and $h^\pm_0=h^\pm$ and $\kappa^\pm_0=\kappa^\pm$.

Prior to prescribing that identification of points between the boundaries at each $\pert$,
we must also prescribe the identification of points amongst each
of the two $\pert$-families of boundaries $\family\Sigma^+_\pert$ and $\family\Sigma^-_\pert$.
After the identification of points, and thus the construction of $\family\Sigma_\pert$,
we are only left with a prescription of the  identification of points along
the $\pert$-family of hypersurfaces $\family\Sigma_\pert$, namely $\hgauge_\pert:\Sigma\to \family\Sigma_\pert$. This gives rise to the so-called hypersurface gauge \cite{Mars:2005}
(see also \cite{mukohyama2000} for a different approach to first
order).
As the families of metrics $\family g^\pm_\pert$ are pulled back onto $\manifold$
using the spacetime gauges at each side to obtain the families of metrics
$g^\pm_\pert=\gauge^{\pm *}_\pert(\family g^\pm_\pert)$ at each side of $\manifold$,
the matching conditions are pulled back using $\hgauge$ onto $\Sigma$
to obtain the relations
\begin{equation}\label{match_pert_fam}
  h^+_\pert=h^-_\pert, \quad \kappa^+_\pert=\kappa^-_\pert,
\end{equation}
where $h^\pm_\pert:=\hgauge^*_\pert(\family h_\pert^\pm)$ and
$\kappa^\pm_\pert:=\hgauge^*_\pert(\family \kappa_\pert^\pm)$.

\begin{figure*}
\centering
  \includegraphics[width=\textwidth]{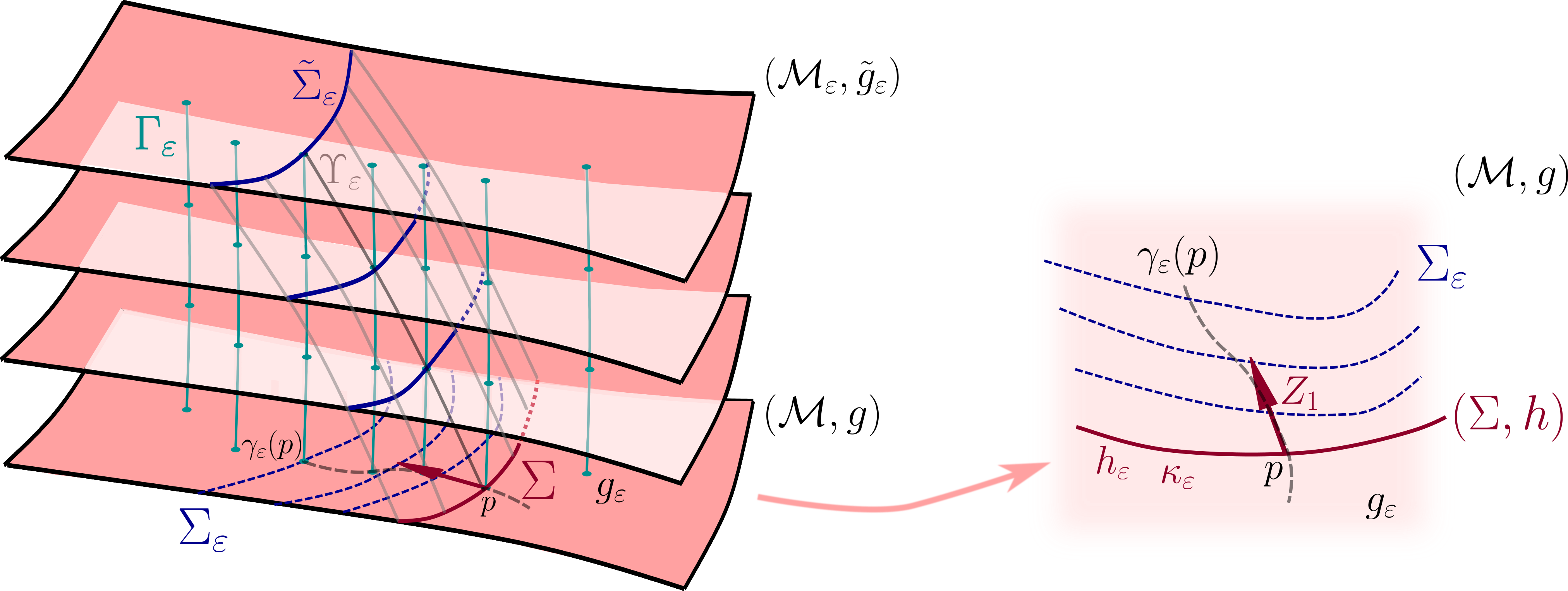}
  \caption{\label{fig:milfulles} Diagram to describe the setup of the perturbation theory for spacetimes and hypersurfaces as described in the main text. This picture applies to both the
  $+$ and $-$ families. The region that is
  to be matched, say $\manifold^+$ with boundary
  $\Sigma^+$, lies either on the left- or on the right-
  hand side of $\Sigma^+$. Observe that we have depicted only
  the members of the family $\manifold_\pert$ for $\pert\geq 0$,
  with $\manifold=\manifold_0$, 
  but we could continue for negative values of $\pert$.
  This is irrelevant because the perturbation procedure only involves the derivatives of the  various $\pert$-family objects evaluated at $\pert=0$, and
  the limit taken from positive $\pert$ equals the limit
taken from negative values by construction.
As shown, $Z_1$ is the tangent
vector of the curve defined by $\gamma_\pert(p)$ at $p$, while
$Z_2$ corresponds to the acceleration of that curve at $p$, and
it is not depicted here.}
\end{figure*}

To understand how the perturbation of the hypersurface is described in this
setting at each side, we simply have to define the family of hypersurfaces $\Sigma^+_\pert$ on
$\manifold^+$ by $\Sigma_\pert^+=\gauge^{-1}_\pert(\family\Sigma_\pert^+)$,
with $\Sigma_0^+=\Sigma^+$ by construction, and the same for the
$-$ side. Let us focus on the $+$ side, the $-$ side will be analogous.
The family of hypersurfaces $\Sigma^+_\pert$ describes how $\Sigma(=\Sigma^+)$ changes
as a set of points in $\manifold^+$. On the other hand, if we take $p\in\Sigma$,
the family of maps $\gamma^+_\pert(p):=\gauge^{+ -1}_\pert(\hgauge^+_\pert(p))$ generates
a curve on $\manifold^+$ starting at $p$ and moving across each $\Sigma^+_\pert$.
The vector field $Z_1$ defined at every point $p$ in $\Sigma^+$ as the velocity of the
curve $\gamma_\pert$, and the acceleration $Z_2$ at $p$ can be decomposed
as $Z^+_1=Q^+_1 \normal^+ + T^+_1$ and $Z^+_2=Q^+_2 \normal^+ + T^+_2$,
for some functions $Q^+_{1/2}$ and tangential vectors $T^+_{1/2}$ to $\Sigma^+$, where
$\normal^+$ is the unit normal to $\Sigma^+$. The information of the deformation
of $\Sigma$, as a set of points, is thus encoded in the functions $Q_1$ and $Q_2$
at first and second order, respectively.
The vectors $T^+_{1/2}$ determine how the points are identified
within the family $\Sigma_\pert$ and therefore
depend on both spacetime and hypersurface gauges.
On the other hand, since the hypersurface
gauge does not modify the matching hypersurfaces as sets of points and only affects how they
are identified pointwise, $Q^+_1$ does
not depend on the hypersurface gauge.
However, at second order both gauges  
get involved in the quantity $Q_2^+$.
This whole construction (dropping the $\pm$ indicators)
is depicted in Fig.~\ref{fig:milfulles}.

The first and second order matching conditions are the first and second
order derivatives of the equations \eqref{match_pert_fam} with respect
to $\pert$ on $\pert=0$. The explicit expressions at each side $\pm$ in terms
of the perturbation tensors $K_1$ and $K_2$, plus $Q_{1/2}$ and $T_{1/2}$
were found in \cite{Mars:2005}. 
The particularization to stationary and axisymmetric
perturbations around a spherical static background
assuming axisymmetric surface deformations was presented in \cite{ReinaVera2015},
while for arbitrarily deformed surfaces the job was done in \cite{MRV2}.
Let us stress that 
this set of perturbed matching conditions arises by demanding that the Riemann tensor does not present a delta distribution,
so it is thus \emph{purely geometric and therefore
  independent of the field equations.}
In \cite{Aranguren:2022} we used those results to write down
the perturbed matching conditions for the two-fluid model at the boundary
of the star, and used them to solve the isolated rotating star global problem.
We will recall the relevant results below, and use
them to obtain the perturbed matching for the tidal problem
for two-fluid stars.

\section{Background}\label{sec:background}

As explained above,
the background configuration is a globally static and spherically symmetric spacetime
composed of the interior and exterior regions of the star.
The geometric configuration is shared by both the isolated rotating star and the tidal problems.
It therefore corresponds to the background configuration constructed in
\cite{comer1999, Andersson_Comer_2001, Aranguren:2022}.
We briefly review the construction of the configuration in this section 
to fix some notation.

We consider two static spherically symmetric spacetimes
with boundary $(\manifold^+,\gback^+,\Sigma^+)$ and $(\manifold^-,\gback^-,\Sigma^-)$ describing the interior and exterior of the star.
In spherical coordinates $\{t_+,r_+,\theta_+,\phi_+\}$
and $\{t_-,r_-,\theta_-,\phi_-\}$ for the corresponding region,
we take
\begin{align*}
  \gback^\pm\!=\!-e^{\nu^\pm(r_\pm)}\diff t_\pm^2+e^{\lambda^\pm(r_\pm)}\diff r_\pm^2+r_\pm^2
  (\diff \theta_\pm^2\!+\!\sin^2\theta_\pm \diff \phi_\pm^2),
\end{align*}
for some pair of functions on each region, $\lambda^\pm$ and $\nu^\pm$.
The boundaries, assumed to be timelike and taken to preserve the spherical and static symmetry \cite{Vera2002},
are given by $\Sigma^\pm:=\{r_\pm=\rstar_\pm\}$, for some positive numbers $\rstar_\pm>0$.
The gluing of $\Sigma^+$ and $\Sigma^-$ is specified, without loss of generality,
by $\theta_+=\theta_-$, $\phi_+=\phi_-$ and $t_+=t_-$ on the boundaries, that we will
denote as $\vartheta$, $\varphi$ and $\tau$, respectively, as coordinates on $\Sigma$.

The interior of the neutron star is described in the background configuration
by the two-fluid model introduced in Sec. \ref{sec:two-fluid}
as described in Sec. \ref{sec:perturbationscheme}.
We are thus given a
master function $\Lambda:=\Lambda(n_0^2,p_0^2,x_0^2)$ as a function
of three arguments. From that we can compute \eqref{eq:ABC}
with $\pert=0$ and construct the quantities
\begin{align*}
  \A^0_0:=&\A_0+2\frac{\partial\B_0}{\partial p_0^2}n_0p_0
    +2\frac{\partial\A_0}{\partial n_0^2}n_0^2
    +2\frac{\partial\A_0}{\partial p_0^2}p_0^2\\
    &+\frac{\partial\A_0}{\partial x_0^2}n_0p_0,\\
  \B^0_0:=&\B_0+2\frac{\partial\B_0}{\partial n_0^2}n_0^2
    +4\frac{\partial\A_0}{\partial n_0^2}n_0p_0
    +\frac{\partial\A_0}{\partial x_0^2}p_0^2,\\
  \C^0_0:=&\C_0+2\frac{\partial\C_0}{\partial p_0^2}p_0^2
    +4\frac{\partial\A_0}{\partial p_0^2}n_0p_0
    +\frac{\partial\A_0}{\partial x_0^2}n_0^2,
\end{align*}
that encode second derivatives.

By definition we first have that $x_0^2=n_0 p_0$.
The interior problem is then composed of
a system of four differential equations for the
set $\{\lambda^+,\nu^+,n_0,p_0\}$.
We refer to Sec. IV.A in \cite{Aranguren:2022} for a full account and explicit expressions
of the background interior problem.
It is convenient to define the mass function in the interior of the star 
as $M^+(r_+)=r_+(1-e^{-\lambda^+(r_+)})/2$.
The exterior solution is given by $e^{\nu^-(r_-)}=e^{-\lambda^-(r_-)}=1-2M/r_-$,
i.e. it is Schwarzschild of mass $M$.

The matching conditions are $\rstar:=\rstar_+=\rstar_-$,
together with $[\lambda]=[\nu]=[\nu^\prime]=0$,
where the prime denotes the derivative with respect to the argument.
Given the field equations, after imposing $\rstar_+=\rstar_-$,
the set of two matching conditions $\{[\lambda]=0,[\nu^\prime]=0\}$
are equivalent to $\Psi_0(\rstar)=0$ and $M=M^+(\rstar)$.
In particular we then have
$e^{\nu(\rstar)}=e^{-\lambda(\rstar)}=1-2M/\rstar$
and
\begin{equation}\label{eq:nuprime}
\nu'(\rstar)=2e^{\lambda(\rstar)}\frac{M}{\rstar^2}=\frac{1}{\rstar}\frac{2M}{R-2M}.
\end{equation}
In addition, the background field equations
can be used to obtain  \cite{Aranguren:2022}
\begin{align}
    &[\lambda^{\prime}]=-\varkappa \rstar e^{\lambda(\rstar)} \Lambda_0(\rstar),\label{eq:dlambda_S}\\
    &[\nu^{\prime\prime}]=-\varkappa
      \left(1+\frac{\rstar\nu^{\prime}(\rstar)}{2}\right)
      e^{\lambda(\rstar)}\Lambda_0(\rstar).\label{eq:ddnu_S}
\end{align}

For some specific forms of $\Lambda$,
the global background problem can be solved, i.e. the solution exists
and is unique, given central values $n_0(0)$ and $p_0(0)$ within some ranges,
at least numerically.

Once the interior problem is solved for some given values of $n_0$ and $p_0$
at the origin, $\Psi_0(\rstar)=0$ fixes the value of $\rstar$ and
$M=M^+(\rstar)$ determines $M$.
We will assume from now on that $\rstar>2M$.
The condition $[\nu]=0$ is just used to set the value at the origin $\nu^+(0)$.
Observe that $n_0(\rstar)$, $p_0(\rstar)$, and thus $\Lambda_0(\rstar)$,
take their values from the interior problem,
are not constrained by the matching whatsoever,
and do not necessarily vanish.

Later we will make use of the background functions
$\mu_0:=n_0\B_0+p_0\A_0$ and $\chi_0:=p_0\C_0+n_0\A_0$.

\section{Isolated rotating star}\label{sec:rotating_star}

The complete analysis for rotating stars is reported in our previous article~\cite{Aranguren:2022}. The reader is addressed to this reference for details on the full sets of equations and a complete description of the computational procedure to solve the global problem at each order. Here, for the sake of completeness, we provide a succinct summary of the approach, using the same notation as in \cite{Aranguren:2022}.

\subsection{First order problem}\label{sec:firstorder}
We assume there exists a class
of gauges for which the first order perturbation tensor at both sides
has the form (we drop the $\pm$ indexes)
\begin{align}
\fpt  =&
 -2r^2\, \omega(r) \sin ^2 \theta \diff t \diff \phi \label{fopert_tensor},
\end{align}
for some function $\omega(r)$ of the radial coordinate only
and bounded at the origin\footnote{We take this as an
  assumption. Although it has been extensively argued in the literature that this is an eventual consequence
  of the global problem, to our knowledge, a full proof of the analogous problem
  in the perfect fluid case has only been produced recently in \cite{MRV1,MRV2}.}.
The field equation in the interior
for $\omega_+$ is given by Eq.~(49) in \cite{Aranguren:2022},
while the equation in the exterior, for $\omega_-$, is
the same with a vanishing right-hand side.

Within the class of gauges that keeps $\fpt$ with the form of \eqref{fopert_tensor},
we have two gauge freedoms to set, one at each region $\pm$, that amount
to the addition of a constant to $\omega_\pm$ correspondingly \cite{ReinaVera2015,MRV2}.
The gauge at the exterior can be fixed so that $\omega_-$ vanishes at infinity.
With that choice the solution is given by
$
\omega_-(r)=2J/r^3,
$
for some constant $J$, which accounts eventually for the total angular momentum.

Finally, the gauge in the interior can be fixed so that
the first order matching conditions read \cite{ReinaVera2015,MRV2,Aranguren:2022}
\[
  [\omega]=[\omega^\prime]=0,
\]
while the deformation quantities $Q_1^\pm(\tau,\vartheta,\varphi)$ satisfy
 $[Q_1]=0$, $Q_1[\lambda']=0$, $Q_1[\nu'']=0$.

The angular momentum of the individual fluids, defined in
\cite{Langlois:1998}, are given explicitly by
\cite{Andersson_Comer_2001}
(we drop the $+$ subindex)
\begin{align*}
  J_n = &-\frac{8\pi}{3}\int_0^\rstar\!\! \diff r\, r^4 e^{(\lambda - \nu)/2}\\
  &\times \left(\mu_0 n_0 (\omega_+-\Omega_n) + \A_0 n_0 p_0 (\Omega_n - \Omega_p)\right),\\
    J_p = &- \frac{8\pi}{3}\int_0^\rstar\!\! \diff r\, r^4 e^{(\lambda - \nu)/2}\\
    &\times\left(\chi_0 p_0 (\omega_+-\Omega_p) + \A_0 n_0 p_0 (\Omega_n - \Omega_p)\right).
\end{align*}
The total angular momentum is recovered with $J = J_n + J_p$.
Similarly, the moments of inertia of the individual fluids are given by
$I_n = J_n/\Omega_n$ and $I_p=J_p/\Omega_p$, and
the total moment of inertia is given by
$I = I_n + I_p$.

\subsection{Second order problem}\label{sec:secondorder}
At second order we assume that there exists a class of gauges in which
the second order perturbation tensor at both sides (dropping the $\pm$ indexes) is given by
the usual form
\begin{align}
\spt = &\left(-4 e^{\nu(r)} h(r, \theta) + 2 r^2{\omega}^2(r) \sin^2\theta\right)\diff t^2 \label{sopert_tensor}\\
& + 4e^{\lambda(r)} v(r, \theta) \diff r^2 +4 r^2 k(r, \theta)
(\diff \theta^2 + \sin ^2 \theta \diff \phi^2),\nonumber
\end{align}
with
\begin{align}
&h(r,\theta)=h_0(r)+h_2(r)P_2(\cos\theta),\nonumber\\
&v(r,\theta)=v_0(r)+v_2(r)P_2(\cos\theta),\label{eq:hvk_legendre}\\
&k(r,\theta)=k_2(r)P_2(\cos\theta),\nonumber
\end{align}
where $P_2(\cos\theta)$ is the Legendre polynomial $P_\ell(\cos\theta)$ with $\ell=2$, and such that all functions are bounded at the origin.
The fact that there is no $k_0(r)$ term fixes partially the class
of gauges in the perturbation scheme.
The gauge freedom that keeps the form \eqref{sopert_tensor}
(see Proposition 6.11 in \cite{MRV1}) together with
\eqref{eq:hvk_legendre} is given by the second order gauge vector
$V_2\propto t\partial_t$ (plus any Killing vector of the background metric $g$).

As for the matter content,
the contribution at second order of the number density of neutrons and protons is
assumed to be of the form
$\eta(r,\theta)=\eta_0(r) + \eta_2(r)P_2(\cos\theta)$ and $\Phi(r,\theta)=\Phi_0(r) + \Phi_2(r)P_2(\cos\theta)$, respectively.
As explained in \cite{Aranguren:2022} in more length,
the fact that there appear no $\ell>2$ terms in the expansions of these quantities
is justified in \cite{Andersson_Comer_2001} using the arguments in the literature
for the perfect fluid problem and assuming equatorial symmetry.

For convenience,
we substitute
the set $\{\eta_\ell(r),\Phi_\ell(r)\}$ by some auxiliary functions $\{\press_{\ell n}(r),\press_{\ell p}(r)\}$ [defined by Eq.~(62) in \cite{Aranguren:2022}] 
that are more easily recognizable as ``pressure''-like functions when compared to the perfect fluid case.

\subsubsection{Second order matching}
Let us consider  $\spt^+$ and $\spt^-$
of the form \eqref{sopert_tensor} with no conditions
on $h(r,\theta)$, $v(r,\theta)$ and $k(r,\theta)$,
and assume that the background and first order matching conditions are satisfied
(no field equations used).
The second order matching conditions are satisfied if and only if
there exists a pair of functions $\defor^\pm(\tau,\vartheta,\varphi)$ on $\Sigma$
and free constants $c_0$, $c_1$, $H_0$ and $H_1$ such that
[Eqs.~(5.69)-(5.75) in \cite{MRV2}, see also Proposition 2 \cite{ReinaVera2015}]
  \begin{subequations}\label{common:mc}
    \begin{align}
      & Q_1[\omega'']=0,\label{common:mc:Q1}\\
      &[ \defor]   =  \ro e^{\lambda(\ro)/2} 
        \left (
        2 c_0 +
        (2 c_1 + H_1) \cos \vartheta  \right ),\label{common:mc:defor}
        \end{align}
        \begin{align}
      &[ k] =  c_0 + c_1 \cos \vartheta,
        \label{common:mc:vh} \\
      &[ h ] = 
        \frac{1}{2} \left ( H_0 + \ro \nu'(\ro) c_0  \right )
        + \frac{1}{4} \ro \nu'(\ro) \left ( H_1 + 2 c_1 \right ) \cos \vartheta,
        \label{common:mc:h}\\
      &\left [ v - 2 k - r k_{,r} \right ]
        =  \left ( H_1 - \frac{1}{2} e^{\lambda(\ro)} \left ( 2 c_1 + H_1
        \right ) \right ) \cos \vartheta
        \nonumber \\
        &\quad + \frac{1}{2} \left [ 
        \defor e^{-\lambda/2} \left ( \frac{\lambda'}{2} - \frac{1}{r}
        \right ) \right ]
        - \frac{1}{4} e^{- \lambda(\ro)} \Qone^2 \left [ \lambda'' \right ], 
        \label{common:mc:q}\\
      &[ h_{,r}] - \frac{\ro \nu'(\ro)}{2}
        [ k_{,r}]
        - \nu'(\ro) \left ( 1 - \frac{\ro \nu'(\ro)}{2} \right )
        [ k] 
        =  \nonumber\\
       &\quad  \frac{\nu'(\ro)}{2}\!
        \left\{\! \left ( 1 - \frac{\ro \nu'(\ro)}{2} \right )
        H_1  - \frac{1}{2} e^{\lambda(\ro)} \left ( 2 c_1 + H_1 \right )\!\right\}
        \cos \vartheta\nonumber\\
       & \quad+ \frac{1}{4} \left [ \defor e^{- \lambda/2}
        \left ( \nu'' + \nu'{}^2 - \frac{\nu'}{r}
        \right ) \right ]
        - \frac{1}{4} e^{-\lambda(\ro)} \Qone^2 
        \left [ \nu''' \right ],
        \label{common:mc:hprime} 
    \end{align}
  \end{subequations}
are satisfied.
The function $\defor^-$ provides the second order deformation,
as seen from the exterior,
since the hypersurface gauge can be partially chosen
so that $Q_2^-=\defor^-$
\cite{ReinaVera2015,MRV2} (see also \cite{Aranguren:2022}).
We have included the full set of second order matching conditions
because we will use them for the tidal problem below.

Now, returning  to the rotating isolated star model,
let us assume the functions $h$, $v$, and $k$ satisfy \eqref{eq:hvk_legendre}.
Then, we necessarily have $c_0=c_1=H_1=0$, cf. \eqref{common:mc:vh}-\eqref{common:mc:h},
and therefore \eqref{common:mc:defor}
yields $[\defor]=0$, so only one function $\defor$
(out of $\defor^\pm$) appears in the matching.
Now, using the decompositions
\begin{align}
&(Q_1)^2(\tau,\vartheta,\varphi)=\sum_{\ell=0}^2 \mathcal{Q}_{\ell}(\tau,\varphi)P_\ell(\cos\vartheta)+\mathcal{Q}_{\perp}(\tau,\vartheta,\varphi),\nonumber\\
  &\defor(\tau,\vartheta,\varphi)=\sum_{\ell=0}^2\defor_{\ell}(\tau,\varphi)P_\ell(\cos\theta)+\defor_{\perp}(\tau,\vartheta,\varphi),\label{eq:Chi_ells}
\end{align}
where we denote by $f_\perp$ the part of $f$ orthogonal to
$\ell=0,1,2$,
and given that the background and first order matching conditions hold,
the set of equations in \eqref{common:mc} is equivalent to the set
$Q_1[\omega'']=0$ plus 
\begin{subequations}\label{mc:l0}
  \begin{align}
    [h_0]=&\frac{1}{2}H_0,\label{match:h0}\\
    [v_0]=&\frac{1}{4} e^{-\lambda(\rstar)/2}\defor_{0}[\lambda']-\frac{1}{4}e^{-\lambda(\rstar)}\mathcal{Q}_{0}[\lambda''],\label{match:v0}\\
    [h'_0]=&\frac{1}{4}e^{-\lambda(\rstar)/2}\defor_{0}[\nu'']-\frac{1}{4}e^{-\lambda(\rstar)}\mathcal{Q}_{0}[\nu'''],\label{match:h0p}
  \end{align}
\end{subequations}
\begin{subequations}\label{mc:l2}
  \begin{align}
  &[k_2]=0,\quad [h_2]=0,\label{mc:k2h2}\\
    &[v_2]-\rstar[k_2']=\frac{1}{4} e^{-\lambda(\rstar)/2}\defor_{2}[\lambda']
            -\frac{1}{4}e^{-\lambda(\rstar)}\mathcal{Q}_{2}[\lambda''],\label{match:k2h2v2} \\
    &[h'_2]-\frac{\rstar}{2}\nu'(\rstar)[k'_2]=
      \frac{1}{4}e^{-\lambda(\rstar)/2}\defor_{2}[\nu'']
      -\frac{1}{4}e^{-\lambda(\rstar)}\mathcal{Q}_2[\nu'''],\label{match:h2p}
  \end{align}
\end{subequations}
and 
\begin{subequations}\label{mc:defor_Q}
\begin{align}
&[\lambda']\, \defor_{1}=[\lambda']\, \defor_{\perp}=0,\quad [\nu'']\, \defor_{1}=[\nu'']\, \defor_{\perp}=0,\\
&[\lambda''] \mathcal{Q}_{1}=[\lambda''] \mathcal{Q}_{\perp}=0,\quad [\nu'''] \mathcal{Q}_{1}=[\nu'''] \mathcal{Q}_{\perp}=0.
\end{align}
\end{subequations}
These last equations for $\defor_1$, $\mathcal{Q}_1$,
$\defor_\perp$, and $\mathcal{Q}_\perp$ are not matching conditions as such,
since their purpose is to determine those quantities involved in the deformation
(in the class of gauges we are working on). Observe that
$\defor_1=\mathcal{Q}_1=\defor_\perp=\mathcal{Q}_\perp=0$ satisfy the relations.

The above analysis of the matching has not taken into account the field equations
at any order (not even the background). If the background
field equations are used, Eqs. \eqref{mc:l0}-\eqref{mc:l2}
take the form of Eqs.~\{(79)-(82)\} in \cite{Aranguren:2022}\footnote{Eq.~(82) in \cite{Aranguren:2022} contains a typo: the second $\defor_2$ should read $\mathcal{Q}_2$.}.
Moreover, \eqref{mc:defor_Q} reduce to
$\Lambda_0(\rstar)\defor_{1}=\Lambda_0(\rstar)\defor_{\perp}=0$
and $\Lambda'_0(\rstar)\mathcal{Q}_{1}=\Lambda'_0(\rstar)\mathcal{Q}_{\perp}=0$.
In any case, if the first order equation for $\omega$ [Eq.~(49) in \cite{Aranguren:2022}]
is also used, then $Q_1[\omega'']=0$ holds automatically.

The global problem, that is, the interior and exterior problems
with common boundary data provided by the matching conditions,
can be split onto the $\ell=0$ and $\ell=2$ sectors. We review
the problems as presented in \cite{Aranguren:2022} next.

\subsubsection{$\ell=0$}
The $\ell=0$ interior problem for the set of
functions $\{h_0^+, v_0^+,\press_{0n},\press_{0p}\}$
comprises Eqs.~\{(65),(67)-(68)\} in \cite{Aranguren:2022}. The exterior solution is given by Eq.~(72) in \cite{Aranguren:2022} after fixing the gauge
  at the exterior so that $h_0^-$ vanishes at infinity (using
  the freedom driven by $V_2^-\propto t\partial_t$ appropriately).
This fixes the spacetime gauge at the exterior completely.

Regarding the matching,
let us first note that
using $V^+_2=H_0 t\partial_t$ at the interior
we can set $H_0=0$ in \eqref{match:h0}. This fixes the spacetime gauge at the interior completely.

As detailed in \cite{Aranguren:2022},
next we must consider the difference of the field equations at both sides
on $\Sigma$. The difference of Eq.~(67) in \cite{Aranguren:2022}
does not provide useful information (just gives $[v_0'])$, but the difference of Eq.~(68) in \cite{Aranguren:2022}
provides, after using the matching up to first order, a relation between
$[h_0']$, $[v_0]$ and $\press_0(\rstar):=n_0\press_{0 n}(\rstar)+p_0\press_{0 p}(\rstar)$. Now, the system composed by
that relation and the two Eqs. \eqref{match:v0} and \eqref{match:h0p}
is shown to be equivalent to one equation for a combination of $\defor_0$ and $\mathcal{Q}_0$,
plus an equation for $[v_0]$ [see \eqref{eq:matching_v0} below], in terms of $\press_0(\rstar)$, and
the original relation from the field equations.

To sum up, the $\ell=0$ sector of the second order perturbative problems match if and only if the two equations
\begin{equation}
  [h_0]=0,\label{eq:matching_h0}
  \end{equation}
  \begin{align}
  &[v_0]=\varkappa\frac{\rstar }{\nu^{\prime}(\rstar)}e^{\lambda(\rstar)}  \label{eq:matching_v0}\\
      &\times\bigg\{\frac{1}{3}\rstar^2 e^{\lambda(\rstar)}n_0(\rstar) p_0(\rstar)\A_0(\rstar)(\Omega_n-\Omega_p)^2
      -\press_0(\rstar)
      \bigg\},\nonumber
\end{align}
hold. Moreover, the matching produces an equation for a combination of $\defor_0$ and $\mathcal{Q}_0$
[Eq.~(91) in \cite{Aranguren:2022}] which we do not include here for brevity.
The field equations produce then a value for $[h_0']$ that is
consistent with the geometrical matching condition \eqref{match:h0p}.

The exterior solution is $h_0^-(r)=-v_0^-(r)$ with
\begin{equation}\label{eq:v0_ext}
v^-_0(r)=\frac{\delta M}{r-2M}-\frac{J^2}{r^3(r-2M)},
\end{equation}
for some constant $\delta M$. Thus, the
$\ell=0$ exterior solution only involves
$\delta M$, which turns out to be the contribution to the mass at second order.
Indeed, the ADM mass
of the family of geometries given by $g_\pert$
at $r\to\infty$, given
some central values $n_0(0)$ and $p_0(0)$, is
$$  M_T=M+\pert^2 \delta M.$$

Now, using the identity $v_0^-(\rstar)\equiv v_0^+(\rstar)-[v_0]$
with \eqref{eq:v0_ext}
and \eqref{eq:matching_v0} we obtain
\begin{widetext}
\begin{equation}
  \label{eq:deltaM}
  \delta M
  =\frac{J^2}{\rstar^3}+(\rstar-2M) v^+_0(\rstar)-\varkappa\frac{\rstar(\rstar-2M)}{\nu^{\prime}(\rstar)}e^{\lambda(\rstar)}
      \bigg\{\frac{1}{3}\rstar^2 e^{\lambda(\rstar)}n_0(\rstar) p_0(\rstar)\A_0(\rstar)(\Omega_n-\Omega_p)^2
      -\press_0(\rstar)
      \bigg\}.
\end{equation}

\end{widetext}
This expression of $\delta M$ corrects the
expression $(60)$ in \cite{Andersson_Comer_2001}, that does not contain
the last term.

\subsection{$\ell=2$}
The $\ell=2$ problem in the interior region
involves the set 
$\{h_2, v_2, k_2,\press_{2n},\press_{2p}\}$ that satisfy the five relations in
Eqs.~\{(66), (69)-(71)\} in \cite{Aranguren:2022}.

The general exterior solution is given explicitly by Eqs.~(73)-(74) in \cite{Aranguren:2022} for {$h_2^-(r)$, $k_2^-(r)$}, and 
\begin{align}
    v_2^{-}(r)=-h_2^{-}(r)+\frac{6J^2}{r^3},\label{eq:v2_ext}
\end{align}
in terms of a free parameter $\constK$ that is to be fixed\footnote{Expression \eqref{eq:v2_ext}
  corrects a typo in the last term in Eq.~(75) of \cite{Aranguren:2022},
  and also (76) of \cite{ReinaVera2015}, where last term should have a global minus.
  This has no other consequences whatsoever.}.
  On the other hand, the pole structure at the origin
  implies that the general interior solution for $\{h_2^+(r),k_2^+(r),v_2^+(r)\}$
  depends on a free parameter, denoted by $A$ in \cite{Aranguren:2022},
  that multiplies the homogeneous part of the solution.

As for the difference on $\Sigma$
of the field equations,
the set of three equations given by
\eqref{mc:k2h2} plus the difference of
[Eqs.~(69)-(71) in \cite{Aranguren:2022}]
is equivalent to the set $\{\eqref{mc:k2h2},\eqref{match:k2h2v2},\eqref{match:h2p}\}$
plus another equation that determines a combination of $\defor_2$ and $\mathcal{Q}_2$,
explicitly given by Eq.~(93) in \cite{Aranguren:2022}.

In short, 
the $\ell=2$ sector of the second order perturbation problems match
if and only if
\begin{align}
  [h_2]=0,\quad[k_2]=0,\label{eq:dif_h2k2}
\end{align}
and then the matching produces an equation for a combination of $\defor_2$ and $\mathcal{Q}_2$
[Eq.~(93) in \cite{Aranguren:2022}].
The field equations produce then values for $[v_2]$, $[h_2']$ and $[k_2']$
consistent with the geometrical equations \eqref{match:k2h2v2} and \eqref{match:h2p}.

Once the interior problem is integrated in terms of the inhomogenous and homogenous
part of the general solution, the parameters $A$ (from the interior) and $C$ (form the exterior) are fixed
using the two relations \eqref{eq:dif_h2k2}.

The value of  $\constK$ of the exterior solution is related with the quadrupole moment $Q$ by
\begin{equation}
  Q=\frac{8}{5}\constK M^3 + \frac{J^2}{M}.
\end{equation}

\section{The tidal problem}\label{sec:tidal}
We summarize the problem for the linearized analysis of perturbations
for a compact body immersed in a quadrupolar tidal field \cite{Hinderer:2008,Damour:2009vw}.
Given a static and spherically symmetric background,
the even-parity first order perturbation tensor in the Regge-Wheeler gauge 
is given by \cite{regge:1957} (we drop the $\pm$ indexes)
\begin{align}
  \tidalpt = &\sum_{\ell,m}\bigg\{e^{\nu(r)}H_{0 \ell m}(r)\diff t^2
               + e^{\lambda(r)}H_{2 \ell m}(r)\diff r^2\nonumber\\
    &+ r^2 K_{\ell m}(r)(\diff\theta^2 + \sin^2\theta\diff\phi^2)\bigg\}Y_{\ell m}(\theta,\phi),
\end{align}
where $Y_{\ell m}(\theta,\phi)$ are the spherical harmonics.
The equations for each mode $\{\ell,m\}$ decouple, and for
$\{\ell\geq 2,m=0\}$ we have that (we drop the $m=0$ label)
\begin{align}
&H_2{}_\ell=H_0{}_\ell,\label{tidal:H2} \\
&K_\ell{}' = H_0{}_\ell'+ \nu{}' H_0{}_\ell,\label{tidal:Kp}\\
&r^2 \nu{}' H_0{}_\ell' = e^{\lambda} \left (\ell(\ell+1) -2 \right)K_\ell \nonumber\\
&\qquad + \left(r (\lambda{}'+\nu{}') - \left(r \nu{}'\right)^2 -e^{\lambda} \ell(\ell+1) + 2 \right)H_0{}_\ell.\label{tidal:H0p} 
\end{align}
This system is usually written as a single second order ODE for $H_{0\ell}$,
and then $K_{\ell}$ is obtained algebraically from \eqref{tidal:H0p}.

For the interior problem the system \eqref{tidal:Kp}-\eqref{tidal:H0p}
is integrated for each pair $\{H^+_{0\ell},K^+_{0\ell}\}$ from a regular origin
(see e.g. \cite{Thorne1967}).
In the exterior vacuum problem, for which
$e^{\nu^-} = e^{-\lambda^-} = 1-2M/r_-$,
the second order equation for $H_0{}_\ell$ is the general Legendre equation (with $m=2$).
The general solution is thus given by
\begin{align}
    H_{0\ell}^-(r_-)=a_{\ell P}\hat{P}_{\ell}^2\left(\frac{r_-}{M}-1\right)+a_{\ell Q}\hat{Q}_{\ell}^2\left(\frac{r_-}{M}-1\right),\label{eq:tidal_ext}
\end{align}
for some constants $a_{\ell P}$, $a_{\ell Q}$ (to keep the notation of \cite{Damour:2009vw}), with
\begin{align*}
    \hat{P}_{\ell}^2(x):=\left(\frac{2^\ell}{\sqrt{\pi}}\frac{\Gamma(\ell+1/2)}{\Gamma(\ell-1)}\right)^{-1} P_{\ell}^2(x),\\
    \hat{Q}_{\ell}^2(x):=\left(\frac{\sqrt{\pi}}{2^{\ell+1}}\frac{\Gamma(\ell+3)}{\Gamma(\ell+3/2)}\right)^{-1} Q_{\ell}^2(x),
\end{align*}
where $P_{\ell}^2(x)$ and $Q_{\ell}^2(x)$ denote the associated Legendre functions of the first and second kind $P_{\ell}^n(x)$ and $Q_{\ell}^n(x)$ with $n=2$, respectively.
The task now is to obtain the necessary and sufficient set of
matching conditions for the interior and exterior problem for the sector
$\{\ell\geq 2,m=0\}$. That constitutes a stationary and axially symmetric perturbation
problem over a static and spherically symmetric background.
Specifically, the problem for a first order
perturbation tensor of the form $\tidalpt$ for $\{\ell\geq 2,m=0\}$ at each side
is equivalent to a second order perturbation problem provided by
\eqref{fopert_tensor} and \eqref{sopert_tensor} with the trivial
choice $\omega=\Omega_n=\Omega_p=0$,
and the substitutions 
\begin{subequations}\label{eq:translate}
\begin{align}
  &h(r,\theta)\to \bar h(r,\theta):=-\frac{1}{4}\sum_{\ell\geq 2}H_{0 \ell}(r,\theta),\\
  &v(r,\theta)\to \bar v(r,\theta):=\frac{1}{4}\sum_{\ell\geq 2} H_{2\ell}(r,\theta),\\
  &k(r,\theta)\to \bar k(r,\theta):=\frac{1}{4}\sum_{\ell\geq 2} K_{\ell}(r,\theta).
\end{align}
\end{subequations}
Since the first order perturbation is vanishing, the second order problem effectively becomes
first order.
Observe that using \eqref{eq:translate}, Eqs.~\{(69), (70), (71)\} in \cite{Aranguren:2022}
with $\omega=f_\omega=0$ translate to Eqs. \eqref{tidal:H2}, \eqref{tidal:Kp}, \eqref{tidal:H0p} for $H_{2\ell}$, $K_{\ell}$ and $H_{0\ell}$ with $\ell=2$, respectively.

The matching conditions for $\tidalpt{}^+$ and
$\tidalpt{}^-$ thus correspond to
\eqref{common:mc} with $\omega=f_\omega=\mathcal{Q}=0$
and the substitutions in \eqref{eq:translate}.
Direct inspection shows that the equations decouple in terms
of $\ell$. Once each $\defor^\pm$ is expanded in Legendre polynomials, see \eqref{eq:Chi_ells},
Eq. \eqref{common:mc:defor} implies that 
$[\defor_\ell]=0$ for $\ell\geq 2$, so we only have one $\defor_\ell$ for each $\ell\geq 2$.
Then, the rest of the matching conditions for $\ell\geq 2$
read
  \begin{align}
    &[K_{\ell}]=0,\quad [H_{0\ell}]=0, \label{mc:tidal1} \\
    &[H_{2\ell}]-\rstar [K_{\ell}']= e^{-\lambda(\rstar)/2}\defor_\ell[\lambda'],\label{mc:tidal2}\\
    &[H'_{0\ell}]+\frac{\rstar}{2} \nu'(\rstar)[K'_\ell]=
      -e^{-\lambda(\rstar)/2}\defor_{\ell}[\nu''].\label{mc:tidal3}
  \end{align}

Now, taking into account the background
field equations, so that \eqref{eq:nuprime},
\eqref{eq:dlambda_S} and \eqref{eq:ddnu_S} hold,
the set of two equations \eqref{mc:tidal2}-\eqref{mc:tidal3}, for each $\ell\geq 2$, is equivalent to the set
  \begin{align}
    &[H_{2\ell}]-\rstar [K_{\ell}']= -\varkappa\rstar e^{\lambda(\rstar)/2}\defor_\ell\Lambda_0(\rstar),\label{mc:tidal2_}\\
    &[H'_{0\ell}]-[K'_\ell]= -\frac{1}{\rstar}
      \left(1+\frac{M}{\rstar}e^{\lambda(\rstar)}\right)
      [H_{2\ell}].\label{mc:tidal3_}
  \end{align}

On the other hand, we must use the information provided by the first order field equations.
Taking the differences of \eqref{tidal:H2}, \eqref{tidal:Kp} and \eqref{tidal:H0p} on $\Sigma$,
and using the background matching conditions and \eqref{mc:tidal1}, together with \eqref{eq:dlambda_S},
we obtain
\begin{align*}
&[H_{2\ell}]=[H_{0\ell}],\quad
  [K'_\ell]=[H'_{0\ell}],\nonumber\\
&[H'_{0\ell}]=-\varkappa \frac{\rstar}{2M}H_{0\ell}(\rstar)\Lambda_0(\rstar),
\end{align*}
respectively. 
The combination of these three equations with
\eqref{mc:tidal1} and \eqref{mc:tidal2_}
is equivalent to the 
set of (six) equations on $\Sigma$ given by
\begin{align}
    &[K_\ell]=0,\quad [H_{0\ell}]=0,\quad [H_{2\ell}]=0,\label{eq:tidalmatching}\\
    &[H_{0\ell}^\prime]=[K_\ell^\prime]=-\varkappa\frac{\rstar^2}{2 M} H_{0\ell}(\rstar)\Lambda_0(\rstar),
           \label{eq:tidalmatchingderivatives}
\end{align}
and
\begin{equation}\label{eq:defor_tidal}
  \left(\frac{\rstar^2}{2M} H_{0\ell}+e^{\lambda(\rstar)/2}\defor_\ell\right)\Lambda_0(\rstar)=0,
\end{equation}
while \eqref{mc:tidal3_} then holds identically.

It is important to note that if the three equations in \eqref{eq:tidalmatching}
hold and the field equations are imposed,
then the two relations in \eqref{eq:tidalmatchingderivatives} are automatically satisfied.
As a result, and to sum up, \emph{the interior and exterior problems for each $\ell$
match if and only if \eqref{eq:tidalmatching} hold,
and then the deformation
satisfies \eqref{eq:defor_tidal}.} The field equations
imply the fulfillment
of the rest of the matching conditions.

\subsection{Computation of the Love numbers}

Given the exterior solution
is provided by \eqref{eq:tidal_ext},
the tidal Love numbers are defined by
$$k_\ell:=\frac{1}{2}\left(\frac{M}{\rstar}\right)^{2\ell+1}a_\ell,$$
where $a_\ell:=a_{\ell Q}/a_{\ell P}$, for each mode $\ell$.
Therefore
\begin{align*}
  a_\ell&=-\frac{\partial_{r_-} \hat{P}_\ell^2-(y_\ell^-/\rstar)\hat{P}_\ell^2}{\partial_{r_-} \hat{Q}_\ell^2-(y_\ell^-/\rstar)\hat{Q}_\ell^2}\bigg\lvert_{r_-=\rstar},
\end{align*}
with $y_\ell:=r H_{0\ell}^\prime/H_{0\ell}$
defined at both $\pm$ sides.

Integrating the interior problem provides
$y_\ell^+(\rstar)$, and then $y_\ell^-(\rstar)$
is obtained using the identity $y_\ell^-(\rstar)=-[y_\ell]+y^+_\ell(\rstar)$.
Equations \eqref{eq:tidalmatching}-\eqref{eq:tidalmatchingderivatives}
establish
\begin{align}\label{eq:jump_y}
    [y_\ell]=-\frac{\varkappa\rstar^3}{2 M}\Lambda_0(\rstar).
\end{align}
As a result,
\begin{align}
    a_\ell&=-\frac{\partial_{r_+} \hat{P}_\ell^2-(y_\ell^+/\rstar)\hat{P}_\ell^2-(\varkappa\rstar^2\Lambda_0(\rstar)/2M)\hat{P}_\ell^2}{\partial_{r_+} \hat{Q}_\ell^2-(y_\ell^+/\rstar)\hat{Q}_\ell^2-(\varkappa\rstar^2\Lambda_0(\rstar)/2M)\hat{Q}_\ell^2}\bigg\lvert_{r_+=\rstar}.\label{eq:a_l}
\end{align}

The error encountered in previous literature (e.g. \cite{Char2018,Yeung:2021})
concerns the
implicit assumption that $y_\ell$ are continuous across the surface of the star,
which is inconsistent with the perturbative procedure if $\Lambda_0(\rstar)$ does not vanish.
Reverting the argument, in those works it was not
proven (not even stated) that
$\Lambda_0(\rstar)=0$ if and only if the functions $y_\ell$ are continuous.

In the numerical analysis we discuss next we will only care about the $\ell=2$ term,
and the tidal problem will be accounted for with the
alternative quantity
$\lambda_2=a_2/3$, known as the tidal deformability. The Love number $k_2$ obtained from
\eqref{eq:a_l}
reads
\begin{align}
  &k_2=\frac{8}{5} \compact^5(1-2\compact)^2(2\compact(\mathcal{Y}-1)-\mathcal{Y}+2)\nonumber\\
  &\times\left\{2\compact\left(4(\mathcal{Y}+1)\compact^4+(6\mathcal{Y}-4)\compact^3+(26-22\mathcal{Y})\compact^2\right.\right.\nonumber\\
  &\left.+3(5\mathcal{Y}-8)\compact-3\mathcal{Y}+6\right)\nonumber\\
     & \left. +3(1-2\compact)^2(2\compact(\mathcal{Y}-1)-\mathcal{Y}+2)\log((1-2\compact))\right\}^{-1}, \label{eq:k2}
\end{align}
where $\compact:=M/R$, with
\begin{equation}
  \mathcal{Y}:=y^-_{\ell=2}(\rstar)=y^+_{\ell=2}(\rstar)+\frac{\varkappa\rstar^3}{2 M}\Lambda_0(\rstar).\label{eq:bigY}
\end{equation}
The form of expression \eqref{eq:k2} coincides (replacing $\mathcal{Y}$ by ``$y$")
with that in the literature,
e.g. \cite{Hinderer2009}, Eq.~(25) in \cite{Char2018}
and Eq.~(24) in \cite{Yeung:2021}. This is
so because ``$y$" there
corresponds to the exterior $y^-_{\ell=2}(\rstar)$, and the exterior problem is the same. However, in \cite{Char2018, Yeung:2021} ``$y$" is given a single value
on the boundary, implicitly assuming \emph{a priori}
that, in the usual wording, $y_{\ell=2}$ is continuous. Because of \eqref{eq:jump_y}, the correct
result requires \eqref{eq:bigY}. The final values
for $k_2$ found in \cite{Char2018, Yeung:2021}
turn out to be valid because, as we have checked,
the EOS forces $\Lambda_0$
to approach zero at the boundary in the models presented in those works.

Expression \eqref{eq:bigY} can be compared with Eq.~(15) in \cite{Hinderer:2010}, which was put forward
following the previous discussion in \cite{Damour:2009vw} for homogeneous stars,
and finally proved for perfect fluids in general in \cite{reina-sanchis-vera-font2017}.
Let us stress, however, that the relation
between the discontinuities of the physical quantities and the jumps of the relevant metric functions is far from obvious, \emph{a priori}. Note that whereas the discontinuities affecting $y_\ell$ and $\delta M$ in the perfect fluid case are both proportional to the value of the energy density at the boundary [see Eqs.~(13) and (24) in \cite{reina-sanchis-vera-font2017}], in the two-fluid model that is not the case. The last term in \eqref{eq:deltaM} is not proportional to $\Lambda_0(\rstar)$.

\section{$I$-Love-$Q$-$\delta M$ relations}
\label{sec:results}

We turn next to apply the theoretical developments of the previous sections to different two-fluid models, spanning three different EOS. Our aim is to investigate if, and by how much, the corrections reported here impact the universality of the $I$-Love-$Q$-$\delta M$ relations, extending our previous work in the perfect fluid with barotropic EOS case~\cite{reina-sanchis-vera-font2017}. For all of our models we impose chemical equilibrium ($\mu_0=\chi_0$).

\subsection{Models}

\subsubsection{Two-fluid polytropic model}
This model is the one presented in \cite{Andersson_Comer_2001}, and describes a star where the individual constituents (the superfluid neutrons and all other components) do not interact, i.e.~there is no entrainment. The corresponding master function is given by
\begin{align}
    \Lambda(n_0^2,p_0^2)=-m_nn_0-\sigma_nn_0^{\beta_n}-m_np_0-\sigma_pp_0^{\beta_p},\label{eq:EoS_AC}
\end{align}
where $m_n$ is the mass of the neutron and
\begin{equation*}
\sigma_n=0.2\,m_n,\quad\beta_n=2.3,\quad\sigma_p=2\,m_n,\quad\beta_p=1.95\,.
\end{equation*}
We choose units such that $m_n=c=G=1$ and the number densities of both fluids ($n_0$ and $p_0$) are given in $\si{fm}^{-3}$
[see 
Eqs.~\eqref{eq:TM_to_meters}-\eqref{eq:TM_to_kilograms} in Appendix \ref{app:units}].

\subsubsection{Toy model}
This EOS is the toy model we suggested in \cite{Aranguren:2022}, for which the master function reads
\begin{equation}
    \Lambda(n_0^2, p_0^2, x_0^2) = -(2 n_0 + p_0 + x_0^2)\,m_n,\label{eq:EoS_TM}
\end{equation}
where the same units as before are chosen. Hence, the same conversion factors to SI units apply 
[Eqs.~\eqref{eq:TM_to_meters}-\eqref{eq:TM_to_kilograms} in Appendix \ref{app:units}]. This equation does account for entrainment between the two fluids, and the energy density at the boundary of the star $\Lambda_0(\rstar)$ turns out to be nonvanishing.

\subsubsection{Relativistic mean field model}
We also consider the model employed in \cite{Yeung:2021} in their study of the $I$-Love-$Q$ relations for superfluid neutron stars. Due to the complexity of the master function and the derived relations, we refer to Appendix \ref{app:functions} for the explicit expressions and the techniques to deal with them within a numerical approach. As shown by Eq.~(\ref{eq:masterfunction}), the master function for this EOS depends on a number of parameters. In our computations we consider the two sets of parameters listed in Table \ref{tab:parameters} for the NL3~\cite{Fattoyev:2010} and GM1~\cite{Glendenning:1991} relativistic mean field models. The conversion factors to SI units are shown in 
Eqs.~\eqref{eq:MFM_to_meters}-\eqref{eq:MFM_to_kilograms} in Appendix \ref{app:units}.

\begin{table}[t!]
\setlength{\tabcolsep}{2.1pt}
\renewcommand{\arraystretch}{1.5}
    \centering
    \begin{tabular*}{\linewidth}{@{\extracolsep{\fill}} c c c c c c }
    \hline\hline
         Model & $c_\sigma^2$ & $c_\omega^2$ & $c_\rho^2$ & $b$ & $c$ \\
    \hline
         NL3 & $15.739$ & $10.530$ & $5.324$ & $0.002055$ & $-0.002650$ \\
         GM1 & $11.785$ & $7.148$ & $4.410$ & $0.002948$ & $-0.001071$ \\
    \hline\hline
    \end{tabular*}
    \caption{Set of parameters for the NL3~\cite{Fattoyev:2010} and GM1~\cite{Glendenning:1991} relativistic mean field models obtained from \cite{Yeung:2021}.}
    \label{tab:parameters}
\end{table}

\subsection{Results}

\begin{figure*}[t]
    \centering
    \includegraphics[width=\textwidth]{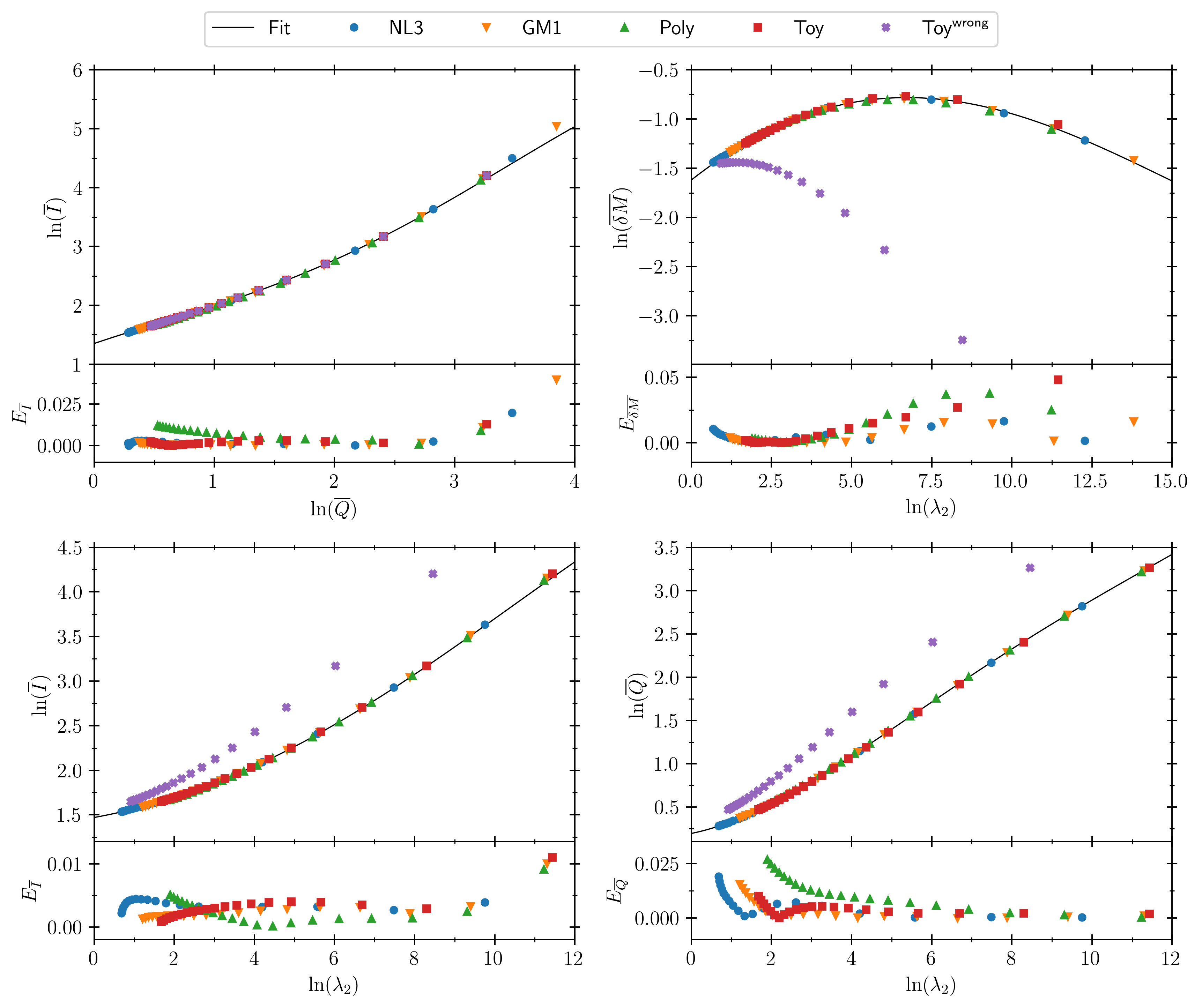}
    \caption{Relations between $\overline{I}-\overline{Q}$ (top-left), $\overline{\delta M}-\lambda_2$ (top-right), $\overline{I}-\lambda_2$ (bottom-left) and $\overline{Q}-\lambda_2$ (bottom-right) for the three models with $\Delta=1$. The purple dots represent the values obtained by taking $[v_0]=0$ in the Toy model, i.e. the incorrect values that would have been predicted with the original HT model. This will only have an effect on the relations involving $\overline{\delta M}$ and $\lambda_2$. The lower panels in each plot represent the relative errors between the individual plots and the fitting curves, $E_X = |(\ln{X}-\ln{X}^{\text{fit}}) / \ln{X}^{\text{fit}}|$. The errors of Toy\textsuperscript{wrong} are not displayed.}
    \label{fig:D1}
\end{figure*}

\begin{figure*}[t]
    \centering
    \includegraphics[width=\textwidth]{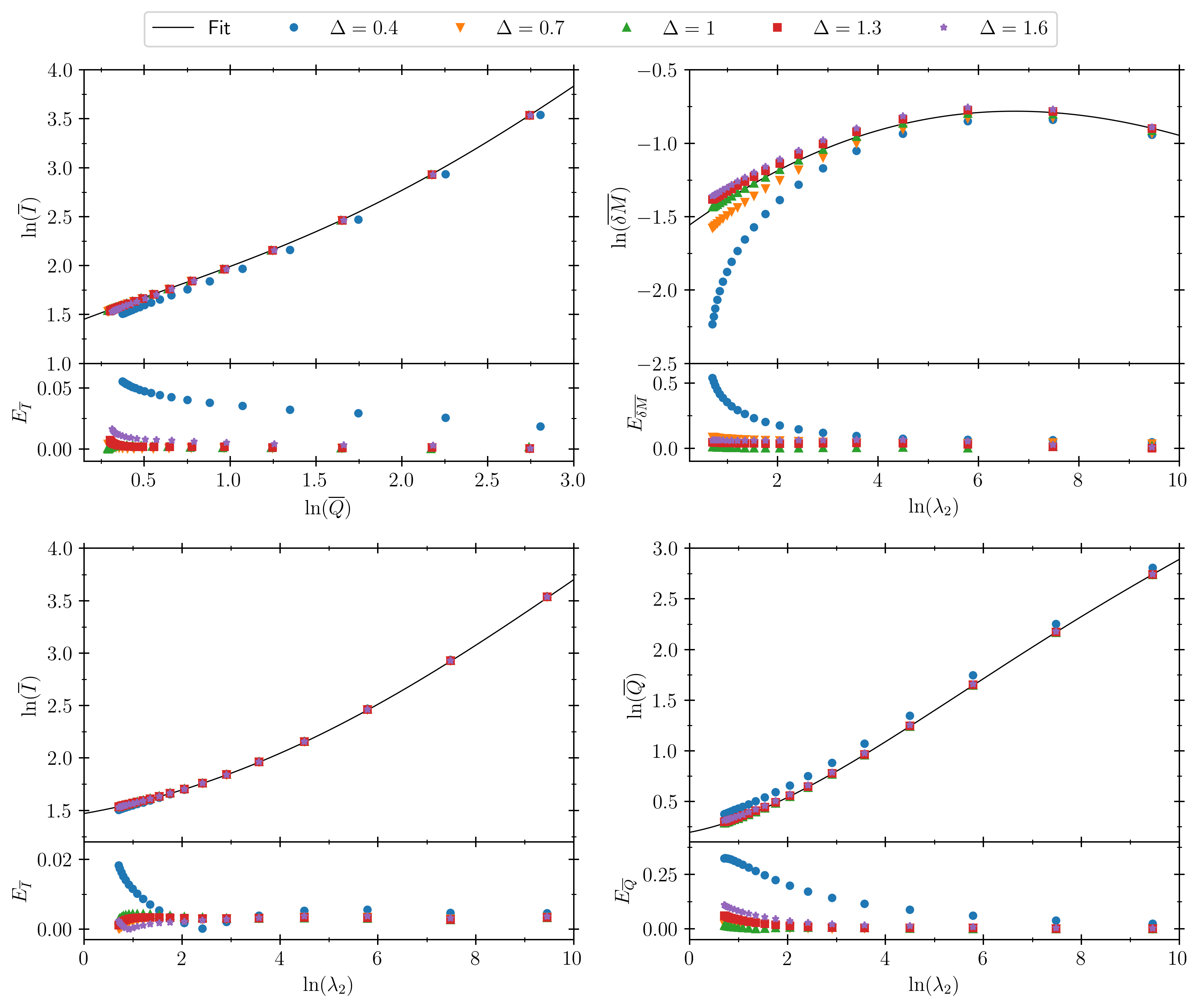}
    \caption{Relations between $\overline{I}-\overline{Q}$ (top-left), $\overline{\delta M}-\lambda_2$ (top-right), $\overline{I}-\lambda_2$ (bottom-left) and $\overline{Q}-\lambda_2$ (bottom-right) for the NL3 model with different values of $\Delta$.}
    \label{fig:NL3}
\end{figure*}

In what follows we show the different $I$-Love-$Q$-$\delta M$ relations we obtain for each of our four models.
The results are displayed in Figs.~\ref{fig:D1} and \ref{fig:NL3}
in terms of the
usual dimensionless
quantities
\begin{align*}
    \overline{I}&:=\frac{I}{M^3}\, , \\
    \overline{Q}&:=\frac{QM}{J^2}\, , \\
    \overline{\delta M}&:=\delta M\frac{M^3}{J^2}\, .
\end{align*}
Note that $M$ refers to the mass of the spherical background configuration.
Although the approximate universal relations involve
only the above dimensionless quantities, we follow the customary practice in the literature and refer to those
relations as the $I$-Love-$Q$-$\delta M$ universal relations (i.e.~not explicitly including the overline in the quantities).

Each of the symbols in Figs.~\ref{fig:D1} and \ref{fig:NL3} corresponds to a particular stellar model, which has been computed numerically using a modification of the code employed in~\cite{reina-sanchis-vera-font2017}. The tidal deformability $\lambda_2$ (which is directly related to the Love number $k_2$) and the contribution to the mass at second order, $\delta M$, are the only quantities that depend on the value of the energy density at the boundary of the star.

Out of the four models we consider here, the toy model is the only one which does present a nonvanishing value of $\Lambda_0(\rstar)$.
For that reason, in the plots in Figs.~\ref{fig:D1} and \ref{fig:NL3} where either $\lambda_2$ or $\overline{\delta M}$ appear, we include an additional set of points labeled  ``Toy\textsuperscript{wrong}''. Those show the results obtained for the same toy model but without considering the correction to the original perturbative frameworks
in which it was (implicitly) assumed all metric functions to be continuous.

In Fig.~\ref{fig:D1} we show the different $I$-Love-$Q$-$\delta M$ relations for the four EOS with $\Delta=1$. The top part of each panel shows the actual correlations between pairs of parameters while the bottom part displays the corresponding relative errors. In all cases the relations found between pairs of parameters follow approximate universal relations (when using the correct
matching conditions). These relations can be accurately fitted with polynomial curves, using the logarithm of the parameters as variables. The individual fitting formulae can be summarized with the expression
\begin{eqnarray}
\ln(y_i)&=&a_i+b_i\ln(x_i)+c_i\ln(x_i)^2\nonumber\\
&&+d_i\ln(x_i)^3+e_i\ln(x_i)^4,\label{eq:universalfit}
\end{eqnarray}
where the values of the coefficients are given in Table \ref{tab:values_fitting}. Those fits are displayed with solid lines in the top panels of each plot in Fig.~\ref{fig:D1}. Our results show that an augmented set of
universal relations for the tidal problem in binary systems of superfluid neutron stars, involving the four perturbation
parameters $\overline{I}$, $\lambda_2$, $\overline{Q}$ and also
$\overline{\delta M}$, exists. This result reproduces the previous findings for the tidal problem in the perfect fluid with barotropic EOS case~\cite{reina-sanchis-vera-font2017}.

\begin{table}[h!]
\setlength{\tabcolsep}{2.1pt}
\renewcommand{\arraystretch}{1.5}
    \centering
    \begin{tabular*}{\linewidth}{@{\extracolsep{\fill}} c c c c c c c }
    \hline\hline
         $y_i$ & $x_i$ & $a_i$ & $b_i$ & $c_i$ & $d_i$ & $e_i$ \\
    \hline
         $\overline{I}$ & $\lambda_2$ & $1.47$  & $0.0817$ & $0.0149$ & $2.87\times10^{-4}$ & $-3.64\times10^{-5}$ \\
         $\overline{I}$ & $\overline{Q}$ & $1.35$ & $0.697$ & $-0.143$ & $9.94\times10^{-2}$ & $-1.24\times10^{-2}$ \\
         $\overline{Q}$ & $\lambda_2$ & $0.194$ & $0.0936$ & $0.0474$ & $-4.21\times10^{-3}$ & $1.23\times10^{-4}$ \\
         $\overline{\delta M}$ & $\lambda_2$ & $-1.619$ & $0.255$ & $-0.0195$ & $-1.08\times10^{-4}$ & $1.81\times10^{-5}$ \\
    \hline\hline
    \end{tabular*}
    \caption{Parameters for the fitting curves of Eq.~(\ref{eq:universalfit}). The first three rows have been obtained from \cite{Yagi:2013awa}, whereas the last row is from \cite{reina-sanchis-vera-font2017} after applying a suitable logarithm base conversion.}
    \label{tab:values_fitting}
\end{table}

The deviation from universality stands out when not taking
into account the
correction to the
formalism (see the purple symbols in the $\overline{\delta M}-\lambda_2$, $\overline{I}-\lambda_2$, and $\overline{Q}-\lambda_2$ plots).
The correct formalism
yields fully universal relations for all the EOS considered,
irrespective of the existence of jumps
of the energy density,
as shown by the red symbols.

In Fig.~\ref{fig:NL3} we illustrate the relations for the specific case of the NL3 model, for different values of $\Delta$. In this case,  universality is lost when the two fluids do not corotate (i.e.~$\Delta\neq1$). The smallest departures from universality are found for the $\overline{I} - \lambda_2$ pair (bottom-left panel in Fig.~\ref{fig:NL3}), with the maximum relative error at the 2\% level. These results were already found in \cite{Yeung:2021}, except for the analysis of the second order contribution to the mass $\overline{\delta M}$. In our study, the inclusion of $\overline{\delta M}$ into the set of quantities to analyze shows that the $\overline{\delta M} - \lambda_2$ curve (top-right panel in Fig.~\ref{fig:NL3}) is significantly more sensitive to the variation of the relative rotation rate between the two fluids $\Delta$ than the rest of the relations. Relative errors as large as 50\% are found for $\Delta=0.4$.

\section{Conclusions}\label{sec:conclusions}

In this work we have studied the tidal problem and the resulting $I$-Love-$Q$ approximate universal relations for rotating superfluid neutron stars in the Hartle-Thorne formalism. To do so we have adapted the stationary and axisymmetric perturbation scheme for global stellar models developed in \cite{Aranguren:2022}
to the first order tidal problem in binary systems.
Our approach is based on the geometrical formalism developed in \cite{ReinaVera2015}
and fully generalized in \cite{MRV1, MRV2}.
The outcome provides the expected correction to the computation
of the Love numbers caused by
a nonvanishing energy density at the interior side
of the stellar boundary. Such correction
is analogous to that of the perfect fluid case
found for homogeneous stars in \cite{Damour:2009vw}
and proven in full generality in \cite{reina-sanchis-vera-font2017}.

The analytic formalism has been applied to different two-fluid stellar models built numerically, spanning three different EOS.
On the one hand, in particular, we have checked that the relevant
physical quantities produced by the EOSs used in \cite{Char2018,Yeung:2021}
tend to zero at the boundary,
thus providing firm grounds to those results.
Further, we have shown how the contribution
to the mass at second order $\delta M$ also satisfies universal relations with $I$, Love and $Q$
for all EOS
when the two fluids corotate ($\Delta=1$).
This result is in agreement with the perfect fluid case \cite{reina-sanchis-vera-font2017}. The universal $I$-Love-$Q$ relations are known
to fail when $\Delta\neq 1$, as shown by \cite{Yeung:2021}. 
We have also found that in the numerical stellar models analyzed in this work, the departure from
universality in the relations
involving $\delta M$ are significantly more sensitive
than the rest. 

The results presented in this paper, thus, complete the set of universal relations for rotating superfluid stars,
generalizing our previous findings in the perfect fluid case. 
Using the extended set of universal relations reported in this work in order to improve observational constraints on the supranuclear EOS of neutron stars is an effort worth pursuing next. Our findings in this direction will be reported elsewhere~\cite{Aranguren:deltaM}.\\

\section{Acknowledgments}
The authors would like to thank Sayak Datta for useful comments on the manuscript. Work supported by the Spanish Agencia Estatal de Investigación (Grants No. PID2021-125485NB-C21 and No. PID2021-123226NB-I00 funded by MCIN/AEI and ERDF A way of making Europe), by the Generalitat Valenciana (Prometeo Grant No. CIPROM/2022/49) and by the Basque Government (Grant No. IT1628-22). E. A. is supported by the Basque Governement Grant No. PRE\_2022\_1\_0190.
N. S. G. is supported by the Spanish Ministerio de Universidades, through a María Zambrano grant (Grant No. ZA21-031) with reference UP2021-044, within the European Union-Next Generation EU. Computations have been performed using the free PSL version of REDUCE, Fortran for the numerical part and the following open-source packages: \texttt{NumPy} \cite{2020Natur.585..357H}, \texttt{SciPy} \cite{2020NatMe..17..261V} and \texttt{Matplotlib} \cite{2007CSE.....9...90H}.

\appendix

\section{Units}\label{app:units}
For the two-fluid polytropic and toy model EOSs, the conversion factors from the given code units (CU) to the SI units were derived in \cite{Aranguren:2022}. They read
\begin{align}
    &r^{\text{SI}} = r^{\text{CU}}\times c\,\sqrt{\frac{\si{fm}^3}{G\,m_n}},\label{eq:TM_to_meters}\\
    &t^{\text{SI}}= t^{\text{CU}}\times\sqrt{\frac{\si{fm}^{3}}{G\,m_n}},\label{eq:TM_to_seconds}\\
    &m^{\text{SI}} = m^{\text{CU}}\times c^3\,\sqrt{\frac{\si{fm}^3}{G^3\,m_n}},\label{eq:TM_to_kilograms}
\end{align}
where $G$, $c$ and $m_n$ recover their SI values.

Similarly, for the mean field model EOS the conversion factors are given by
\begin{align}
    &r^{\text{SI}}=r^{\text{CU}}\times c\sqrt{\frac{c}{G\hbar}}\,\si{fm}^2,\label{eq:MFM_to_meters}\\
    &t^{\text{SI}}=t^{\text{CU}}\times \sqrt{\frac{c}{G\hbar}}\,\si{fm}^2,\label{eq:MFM_to_seconds}\\
    &m^{\text{SI}}=m^{\text{CU}}\times c^3\sqrt{\frac{c}{G^3\hbar}}\,\si{fm}^2.\label{eq:MFM_to_kilograms}
\end{align}
Note that $m$ is a general unit of mass, given in $\si{kg}$, not the nucleon mass mentioned in Appendix \ref{app:functions}.

\begin{widetext}
\section{Mean field model EOS}\label{app:functions}
The master function for this EOS is given by \cite{Kheto2015}
\begin{align}
\Lambda_0 = & -\frac{\comega}{18\pi^4}(\kn^3+\kp^3)^2 - \frac{\crho}{72\pi^4}(\kp^3-\kn^3)^2-\frac{1}{4\pi^2} \left(\kn^3\sqrt{\kn^2+\mds}+\kp^3\sqrt{\kp^2+\mds}\right)\nonumber\\
&-\frac{1}{4\csigma} \left\{(2m-\ms)(m-\ms)+\ms\left(bm\csigma(m-\ms)^2+c\csigma(m-\ms)^3\right)\right\}\nonumber\\
&-\frac{1}{3}bm(m-\ms)^3-\frac{1}{4}c(m-\ms)^4-\frac{1}{8\pi^2}\left\{\kp(2\kp^2+\me^2)\sqrt{\kp^2+\me^2}-\me^4 \ln\left(\frac{\kp+\sqrt{\kp^2+\me^2}}{\me}\right)\right\},\label{eq:masterfunction}
\end{align}
where $\kn=(3\pi^2n_0)^{1/3}$, $\kp=(3\pi^2p_0)^{1/3}$, $m$ is the nucleon mass (the average of the neutron and proton masses), and the parameter $\ms$ is the Dirac effective mass, coming from the transcendental equation
\begin{align}
\ms= & m-\ms\frac{\csigma}{2\pi^2}\Bigg\{\kn\sqrt{\kn^2+\mds}+\kp\sqrt{\kp^2+\mds}+\frac{1}{2}\mds\ln\left(\frac{-\kn+\sqrt{\kn^2+\mds}}{\kn+\sqrt{\kn^2+\mds}}\right)\nonumber\\
&+\frac{1}{2}\mds\ln\left(\frac{-\kp+\sqrt{\kp^2+\mds}}{\kp+\sqrt{\kp^2+\mds}}\right)\Bigg\}+bm\csigma(m-\ms)^2+c\csigma(m-\ms)^3.\label{eq:transcendental}
\end{align}

For convenience, we may work instead with a differential equation for $\ms$ (this strategy was discussed in \cite{ComerJoynt2003}, although for a different expression of the EOS)
\begin{align*}
m_\star^{\prime}\lvert_0=\frac{\partial m_\star}{\partial\kn}\bigg\lvert_0\kn^{\prime}+\frac{\partial m_\star}{\partial\kp}\bigg\lvert_0\kp^{\prime},
\end{align*}
where (again \cite{Kheto2015})
\begin{align*}
\frac{\partial m_\star}{\partial\kn}\lvert_0=&-\frac{\csigma}{\pi^2}\frac{\ms\kn^2}{\sqrt{\kn^2+\mds}}\Bigg\{\frac{3m-2\ms+3bm\csigma(m-\ms)^2+3c\csigma(m-\ms)^3}{\ms}\nonumber\\
&-\frac{\csigma}{\pi^2}\left(\frac{\kn^3}{\sqrt{\kn^2+\mds}}+\frac{\kp^3}{\sqrt{\kp^2+\mds}}\right)+2bm\csigma(m-\ms)+3c\csigma(m-\ms)^2\Bigg\}^{-1},
\end{align*}
\begin{align*}
\frac{\partial m_\star}{\partial\kp}\bigg\lvert_0=&-\frac{\csigma}{\pi^2}\frac{\ms\kp^2}{\sqrt{\kp^2+\mds}}\Bigg\{\frac{3m-2\ms+3bm\csigma(m-\ms)^2+3c\csigma(m-\ms)^3}{\ms}\nonumber\\
&-\frac{\csigma}{\pi^2}\left(\frac{\kn^3}{\sqrt{\kn^2+\mds}}+\frac{\kp^3}{\sqrt{\kp^2+\mds}}\right)+2bm\csigma(m-\ms)+3c\csigma(m-\ms)^2\Bigg\}^{-1}.
\end{align*}

The generalized pressure is given by
\begin{align*}
\Psi_0=\Lambda_0+\frac{1}{3\pi^2}\left(\mu_0\kn^3+\chi_0\kp^3\right),
\end{align*}
where the two auxiliary functions $\mu_0$ and $\chi_0$
explicitly read
\begin{align*}
&\mu_0=\frac{\comega}{3\pi^2}\left(\kn^3+\kp^3\right)-\frac{\crho}{12\pi^2}\left(\kp^3-\kn^3\right)+\sqrt{\kn^2+\mds},\\
&\chi_0=\frac{\comega}{3\pi^2}\left(\kn^3+\kp^3\right)+\frac{\crho}{12\pi^2}\left(\kp^3-\kn^3\right)+\sqrt{\kp^2+\mds}+\sqrt{\kp^2+\me^2}.
\end{align*}

The functions accounting for the first and second order derivatives of $\Lambda_0$ are given by
\begin{align*}
\A_0=&\comega-\frac{1}{4}\crho+\frac{\comega}{5\mu_0^2}\left\{2\kp^2\frac{\sqrt{\kn^2+\mds}}{\sqrt{\kp^2+\mds}}+\frac{\comega}{3\pi^2}\left(\frac{\kn^2\kp^3}{\sqrt{\kn^2+\mds}}+\frac{\kp^2\kn^3}{\sqrt{\kp^2+\mds}}\right)\right\}\nonumber\\
&+\frac{\crho}{20\mu_0^2}\left\{2\kp^2\frac{\sqrt{\kn^2+\mds}}{\sqrt{\kp^2+\mds}}+\frac{\crho}{12\pi^2}\left(\frac{\kn^2\kp^3}{\sqrt{\kn^2 + \mds}}+\frac{\kp^2\kn^3}{\sqrt{\kp^2+\mds}}\right)\right\}\nonumber\\
&-\frac{\crho\comega}{30\mu_0^2\pi^2}\left(\frac{\kn^2\kp^3}{\sqrt{\kn^2+\mds}}-\frac{\kp^2\kn^3}{\sqrt{\kp^2+\mds}}\right)+\frac{3\pi^2\kp^2}{5\mu_0^2\kn^3}\frac{\kn^2+\mds}{\sqrt{\kp^2+\mds}},
\end{align*}
\begin{align*}
\B_0=&\frac{3\pi^2\mu_0}{\kn^3}-\comega\frac{\kp^3}{\kn^3}+\frac{1}{4}\crho\frac{\kp^3}{\kn^3}-\frac{\comega\kp^3}{5\mu_0^2\kn^3}\left\{2\kp^2\frac{\sqrt{\kn^2+\mds}}{\sqrt{\kp^2+\mds}}+\frac{\comega}{3\pi^2}\left(\frac{\kn^2\kp^3}{\sqrt{\kn^2+\mds}}+\frac{\kp^2\kn^3}{\sqrt{\kp^2+\mds}}\right)\right\}\nonumber\\
&-\frac{\crho\kp^3}{20\mu_0^2\kn^3}\left\{2\kp^2\frac{\sqrt{\kn^2+\mds}}{\sqrt{\kp^2+\mds}}+\frac{\crho}{12\pi^2}\left(\frac{\kn^2\kp^3}{\sqrt{\kn^2+\mds}}+\frac{\kp^2\kn^3}{\sqrt{\kp^2+\mds}}\right)\right\}\nonumber\\
&+\frac{\crho\comega\kp^3}{30\pi^2\mu_0^2\kn^3}\left(\frac{\kn^2\kp^3}{\sqrt{\kn^2+\mds}}-\frac{\kp^2\kn^3}{\sqrt{\kp^2+\mds}}\right)-\frac{3\pi^2\kp^5}{5\mu_0^2\kn^6}\frac{\kn^2+\mds}{\sqrt{\kp^2+\mds}},
\end{align*}
\begin{align*}
\C_0=&\frac{3\pi^2\chi_0}{\kp^3}+\frac{1}{4}\crho\frac{\kn^3}{\kp^3}-\comega\frac{\kn^3}{\kp^3}-\frac{\comega\kn^3}{5\mu_0^2\kp^3}\left\{2\kp^2\frac{\sqrt{\kn^2+\mds}}{\sqrt{\kp^2+\mds}}+\frac{\comega}{3\pi^2}\left(\frac{\kn^2\kp^3}{\sqrt{\kn^2+\mds}}+\frac{\kp^2\kn^3}{\sqrt{\kp^2+\mds}}\right)\right\}\nonumber\\
&-\frac{\crho\kn^3}{20\mu_0^2\kp^3}\left\{2\kp^2\frac{\sqrt{\kn^2+\mds}}{\sqrt{\kp^2+\mds}}+\frac{\crho}{12\pi^2}\left(\frac{\kn^2\kp^3}{\sqrt{\kn^2+\mds}}+\frac{\kp^2\kn^3}{\sqrt{\kp^2+\mds}}\right)\right\}\nonumber\\
&+\frac{\crho\comega\kn^3}{30\pi^2\mu_0^2\kp^3}\left(\frac{\kn^2\kp^3}{\sqrt{\kn^2+\mds}}-\frac{\kp^2\kn^3}{\sqrt{\kp^2+\mds}}\right)-\frac{3\pi^2}{5\mu_0^2\kp}\frac{\kn^2+\mds}{\sqrt{\kp^2+\mds}},
\end{align*}
\end{widetext}
together with
\begin{align*}
\A_0^0=\comega-\frac{\crho}{4}+\frac{\pi^2}{\kp^2}\frac{\ms\frac{\partial m_\star}{\partial\kp}\lvert_0}{\sqrt{\kn^2+\mds}},
\end{align*}
\begin{align*}
\B_0^0=\comega+\frac{\crho}{4}+\frac{\pi^2}{\kn^2}\frac{\kn+\ms\frac{\partial m_\star}{\partial \kn}\lvert_0}{\sqrt{\kn^2+\mds}},
\end{align*}
and
\begin{align*}
\C_0^0&=\comega+\frac{\crho}{4}+\frac{\pi^2}{\kp^2}\frac{\kp+\ms\frac{\partial m_\star}{\partial\kp}\lvert_0}{\sqrt{\kp^2+\mds}}+\frac{\pi^2}{\kp}\frac{1}{\sqrt{\kp^2+\me^2}}.
\end{align*}
\\

\bibliography{super_fluid}
\bibliographystyle{ieeetr}

\end{document}